\documentclass[aps,prd,floatfix,twocolumn,reprint,amsmath,amssymb,superscriptaddress,longbibliography]{revtex4-1}
\usepackage{amsfonts}
\usepackage{mathrsfs}
\usepackage{amsmath}
\usepackage{color}
\usepackage{natbib}
\usepackage{graphicx}
\usepackage{bm}
\usepackage{amssymb}
\usepackage{xspace}
\usepackage{epstopdf}
\usepackage{dcolumn}
\usepackage{multirow}
\usepackage[colorlinks=true, letterpaper=true, pdfstartview=FitV, linkcolor=blue, citecolor=blue, urlcolor=black]{hyperref}

\makeatletter

\newcommand{\Rmnum}[1]{\expandafter\@slowromancap\romannumeral #1@}
\makeatother

\begin{document}


\title{Anomalous spatial shifts in interface electronic scattering}

\author{Zhi-Ming Yu}

\address{Research Laboratory for Quantum Materials, Singapore University of
Technology and Design, Singapore 487372, Singapore}

\author{Ying Liu}

\address{Research Laboratory for Quantum Materials, Singapore University of
Technology and Design, Singapore 487372, Singapore}

\author{Shengyuan A. Yang}


\address{Research Laboratory for Quantum Materials, Singapore University of
Technology and Design, Singapore 487372, Singapore}

\begin{abstract}
The anomalous spatial shifts at interface scattering, first studied in geometric optics, recently found their counterparts in the electronic context. It was shown that both longitudinal and transverse shifts, analogous to the Goos-H\"{a}nchen and Imbert-Fedorov effects in optics, can exist when electrons are scattered at a junction interface. More interestingly, the shifts are also discovered in the process of Andreev reflection at a normal/superconductor interface. Particularly, for the case with unconventional superconductors, it was discovered that the transverse shift can arise solely from the superconducting pair potential and exhibit characteristic features depending on the pairing. Here, we briefly review the recent works in this field, with an emphasis on the physical picture and theoretical understanding.
\end{abstract}
\maketitle

\section{Introduction}

The analogy between electronics and optics has inspired many breakthroughs in both fields. One common phenomenon for both is the presence of scattering when electrons/photons hit an interface. As a fundamental physical process, such interface scattering provides an essential mechanism for modulating the propagation of electrons/photons, which in turn constitutes the foundation for the electronic/optical device design.

\begin{figure}[t]
\includegraphics[width=8cm]{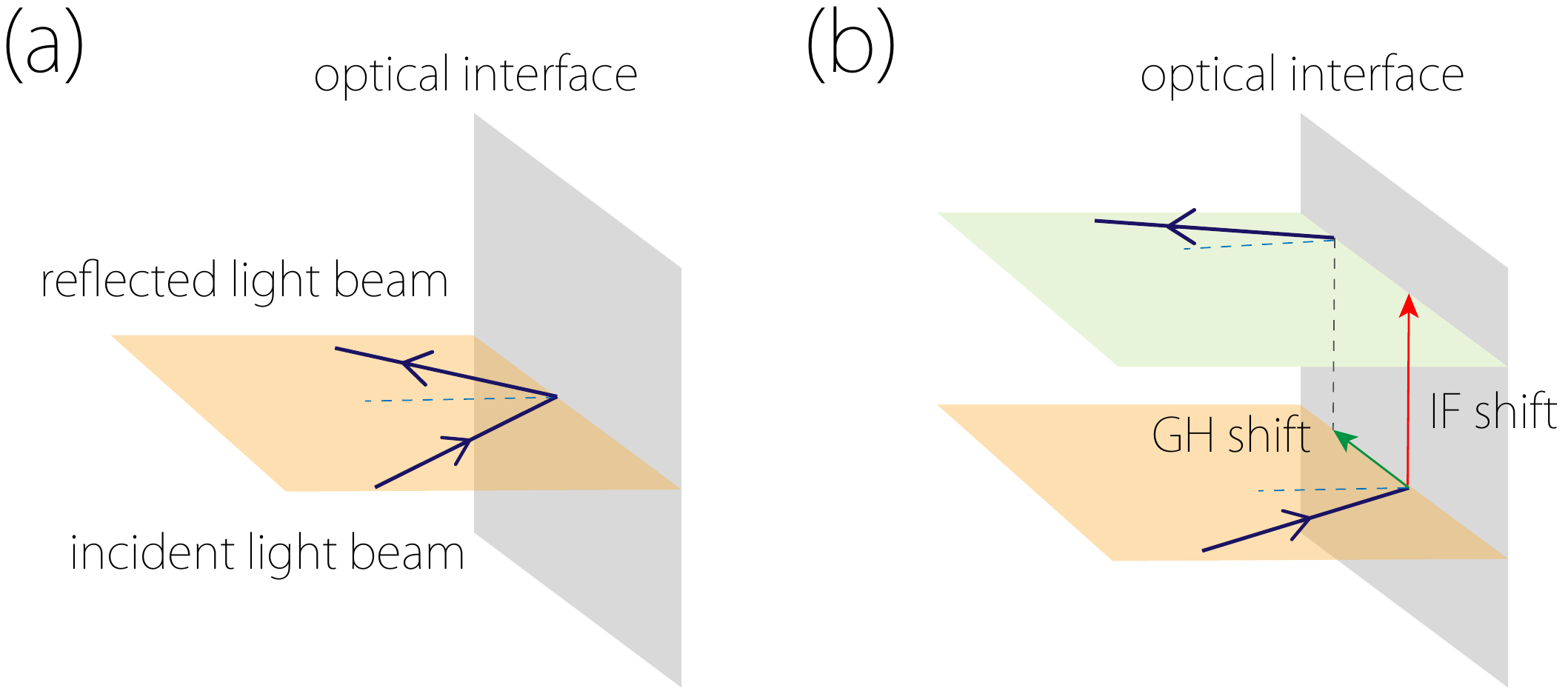}\caption{Illustrations showing (a) the usual picture of reflection for a light beam at an optical interface, obeying the laws of reflection; and (b) in certain cases, the reflected light beam acquires an anomalous spatial shift, including the longitudinal (GH effect) and the transverse (IF effect) components.\label{fig1}}
\end{figure}

Probably the first knowledge we learn about interface scattering is on the reflection of a light beam at a \emph{flat} optical interface (i.e., an interface between different optical media). The effect is summarized by the \emph{laws of reflection} and mathematically described by the Fresnel equations. Specifically, the laws state that: (i) The incident beam, the interface normal, and the reflected beam lie in the same plane; (ii) the incident beam and the reflected beam are on opposite sides of the normal (i.e., the reflection is specular), and they make the same angle with the normal. In addition, for a sharp interface, it is tacitly understood that the incident beam and the reflected beam meet at the same point on the interface. These are illustrated in Fig. \ref{fig1}(a).

These laws have been known since ancient Greek times. However, later studies showed that the they need revision in certain cases. In a work published in 1947 \cite{Goos1947}, Goos and H\"{a}nchen pointed out that when the light beam undergoes a total reflection, the incident and the reflected beams may not meet at the same point on the interface, rather, there generally exists a \emph{longitudinal} spatial shift between them within the plane of incidence [see Fig.~\ref{fig1}(b)]. This shift, known as the Goos-H\"{a}nchen shift, has been studied and verified in many different contexts, and is established as a powerful technique to probe interface properties in optics, acoustics, and atomic physics~\cite{Fornel2010}.

More interestingly, the works by Fedorov (1955) and Imbert (1972) challenged the statement (i) in the laws of reflection \cite{Fedorov1955,Imbert1972}. They found that although the law may hold well for a non-polarized light beam, for a circularly polarized light, however, the plane of reflection (defined by the reflected beam and the normal) may be different from the plane of incidence, i.e., there exists a \emph{transverse} spatial shift for the reflection [see Fig.~\ref{fig1}(b)]. This shift is known as the Imbert-Fedorov shift.

The Goos-H\"{a}nchen shift and the Imbert-Fedorov shift were initially derived based on the Maxwell equations, which are the classical description of electromagnetic waves. Hence, the two effects can be regarded as general wave phenomena. In 2004, Onoda \emph{et al.} offered new insight into these effects \cite{OnodaPRL2004}, by developing semiclassical equations of motion for the light wave packet analogous to the Chang-Sundaram-Niu equations for electrons \cite{ChangPRL1995,ChangPRB1996,SundaramPRB1999}.
The Imbert-Fedorov shift was reproduced from the equations of motion for an interface scattering process. At that time, the electron spin Hall effect was a hot topic~\cite{Murakami2003,Sinova2004}, hence the opposite Imbert-Fedorov shifts for opposite circular polarizations was interpreted as an optical spin Hall effect \cite{OnodaPRL2004}. This rekindled interest in these optical effects in recent years~\cite{Bliokh2015}, and the advance in optical measurement technique enabled quantitative comparisons between theory and experiment~\cite{Hosten2008,Yin2013}.

Since electrons also exhibit wave behavior in propagation, as described by the Schr\"{o}dinger equation, one naturally wonders whether analogies of the optical anomalous shifts exist for electronic interface scattering. In fact, proposals of the longitudinal (Goos-H\"{a}nchen like) shift for electrons appeared quite early, at least since the 1970s~\cite{Miller1972,Fradkin1974a,Fradkin1974b}. Since the early 2000s, the longitudinal shift was actively explored in the two-dimensional (2D) electron gas with spin-orbit coupling (SOC) due to the surge of research interest in spintronics \cite{Sinitsyn2005,Chen2008,ChenPRB2011}, and then in graphene and 2D material heterostructures \cite{BeenakkerGH2009,ZhaoPRB2010,Sharma2011,WuPRL2011,Chen2011,ChenEPJB2013}. The research is strongly motivated by the rapid progress of the experimental techniques in fabricating high-quality junctions and in manipulating electrons in these micro-structures \cite{ChenSci2016}. There emerged a field of electron optics, which targets at the accurate control of electron propagation as in optics~\cite{Spector1990,Molenkamp1990,Dragoman1999}.

On the other hand, the electronic analog of the Imbert-Fedorov shift was revealed only recently. In 2015, Jiang \emph{et al.} \cite{JiangPRL2015} and Yang \emph{et al.} \cite{YangPRL2015}, via different approaches, predicted the presence of the transverse shift for electrons in a special type of 3D materials---the Weyl semimetals. They showed that the Berry curvatures play a key role in the effect, and the shift leads to a chirality Hall effect in a Weyl semimetal junction \cite{YangPRL2015}. Later, the effect was used to explain the high mobility observed for Weyl semimetals, and was extended to the closely related multi-Weyl semimetals \cite{WangPRB2017}. The possible effect of topological Fermi arcs of a Weyl semimetal on the shift was investigated recently \cite{Chattopadhyay2018}.

In all the above-mentioned effects, the incident and the scattered beams are of the same kind of particles, i.e., an electron is scattered as an electron, and a photon is scattered as a photon. Yet there exists a special kind of scattering process occurring at a normal-metal/superconductor (NS) interface---the Andreev reflection \cite{Andreev1964,Gennes1966}, in which an incident electron is reflected back as a hole. Here, the incident and the scattered particles are of different identity, and even the electric charge is changed in scattering. One also notes that the Andreev reflection may also violate the statement (ii) in the laws of reflection: the reflected beam and the incident beam are typically on the same side of the normal, i.e., it is a retroreflection.
With these nontrivial features, it is thus intriguing to ask whether the anomalous shifts also happen in Andreev reflections. This question is answered in the affirmative by Liu, Yu, and Yang in 2017 \cite{LiuSUTDPRB2017}. They showed that the effect generally exists for a heterojunction consisting of 3D metal with SOC and a conventional $s$-wave superconductor. In subsequent works, Liu \emph{et al.} \cite{LiuSUTD2018} analyzed in detail the longitudinal shift for such a system.
Remarkably, Yu \emph{et al.} \cite{YuSUTDPRL2018} showed that a sizable transverse shift could be induced solely by unconventional pairings on the superconductor side, and exhibits characteristic behaviors corresponding to the symmetry of the pair potential. Most recently, the transverse shift was also proposed for the crossed Andreev reflection process \cite{LiuCAR2018}, a nonlocal scattering process for which the incident electron and the scattered hole are at different normal metal terminals connected to a single piece of superconductor \cite{Byers1995,Deutscher2000}.

In this paper, we provide a brief review on these recent advancements in the study of the anomalous shifts in electronic systems. We shall be focusing on the transverse shift (although the longitudinal shift in Andreev reflection will be included). As for the studies on the longitudinal shift in electronic systems, we refer the readers to the previous review paper Ref.~\cite{ChenReview2013}. Excellent reviews on the shifts in the optical context can be found in Ref.~\cite{BliokhReviw}.

\begin{figure}[t]
\includegraphics[width=8cm]{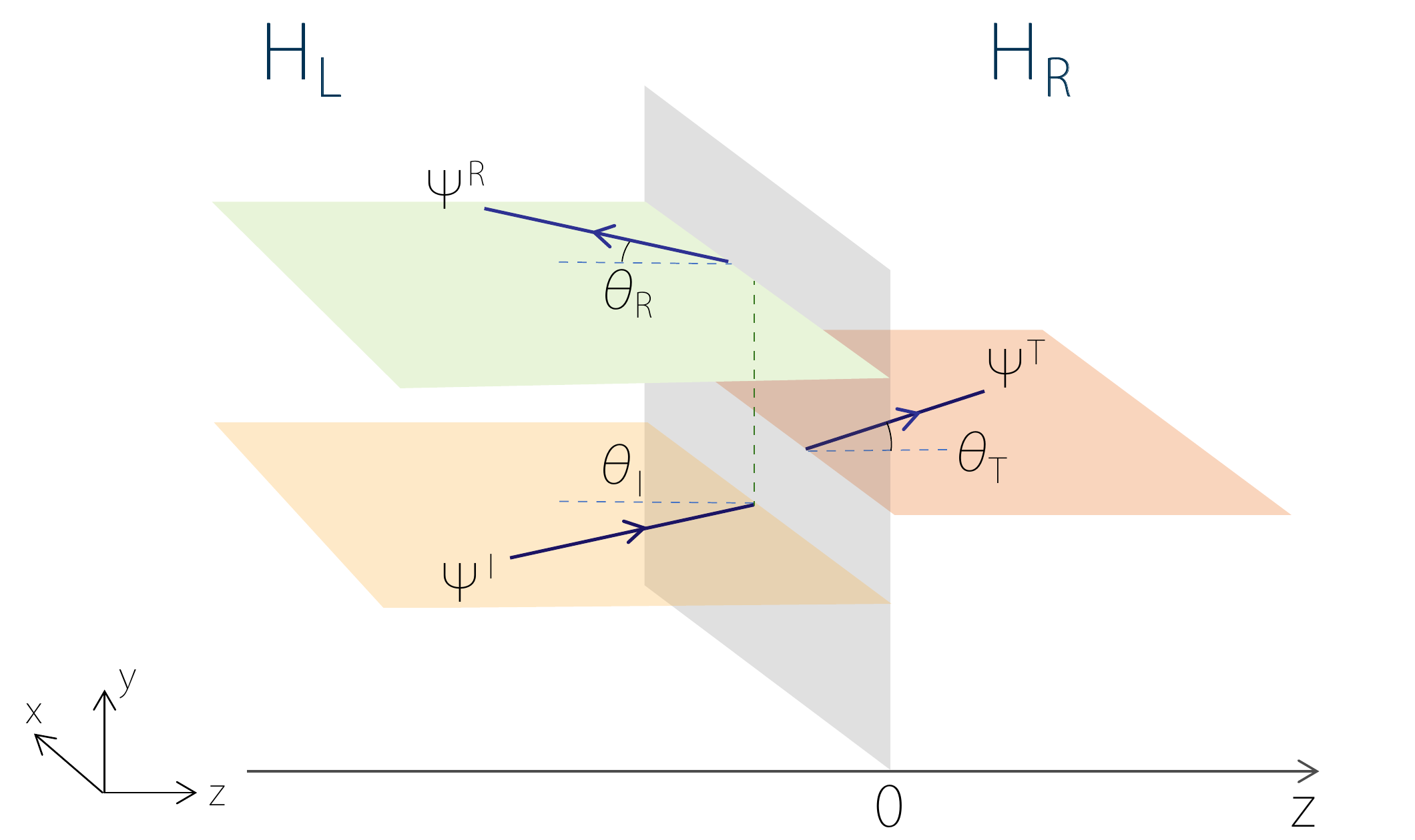}\caption{Schematic figure showing the basic setup for investigating the anomalous spatial shifts in interface scattering.\label{fig2}}
\end{figure}

\section{Basic Setup}

We first discuss the basic setup for investigating the anomalous spatial shifts in interface scattering. Certainly we need an interface between two different media, where the electronic scattering occurs. The interface is assumed to be flat and clean, and without loss of generality, we assume it is located at $z=0$, as illustrated in Fig. \ref{fig2}. The whole system is assumed to be extended along $x$ and $y$ directions, which amounts to saying that the system dimension in these two directions is much larger than the particle wavelength (and also the anomalous shift).

The two media on the two sides of the interface are described by two model Hamiltonians $H_L$ and $H_R$, respectively. An interfacial barrier may be modeled by adding terms such as $h\delta(z)$ to the model. The junction is assumed to be clean (in other words, the system dimension is assumed to be within a mean free path), such that an incident particle does not experience other (disorder) scattering except the scattering at the interface. Due to the translational symmetry in $x$ and $y$, the transverse wave vector $\bm k_\|=(k_x,k_y)$ for the incident particle will be a conserved quantity during scattering.

For systems with a rotational symmetry along $z$, without loss of generality, one can assume the plane of incidence is the $x$-$z$ plane (see Fig. \ref{fig2}). Then the longitudinal shift is in the $x$ direction, and the transverse shift is in the $y$ direction. However, if the system does not possess the rotational symmetry, the result will depend on the orientation of the plane of incidence. This is the case for the systems with anisotropic pair potentials, where one needs to specify a rotation angle $\alpha$ for the incident plane with respect to the crystal axis (see Fig. \ref{fig: ill_junct}).

The anomalous positional shifts are defined for laterally \emph{confined} particle beams. They cannot be defined for the unconfined plane wave states. Hence, one typically assumes an incident particle beam $\bm\Psi^I$ coming from the left media, computes the reflected beam $\bm\Psi^R$ and the transmitted beam $\bm\Psi^T$, and compares their center positions at the interface to obtain the shifts in reflection and in transmission.

The setup described here can be easily extended to include more than one interfaces. For example, in the study of crossed Andreev reflection, one considers a normal/superconductor/normal (NSN) sandwich structure with two interfaces (see Fig.~\ref{fig: ill_CAR}).

\begin{figure}[t]
\includegraphics[width=8cm]{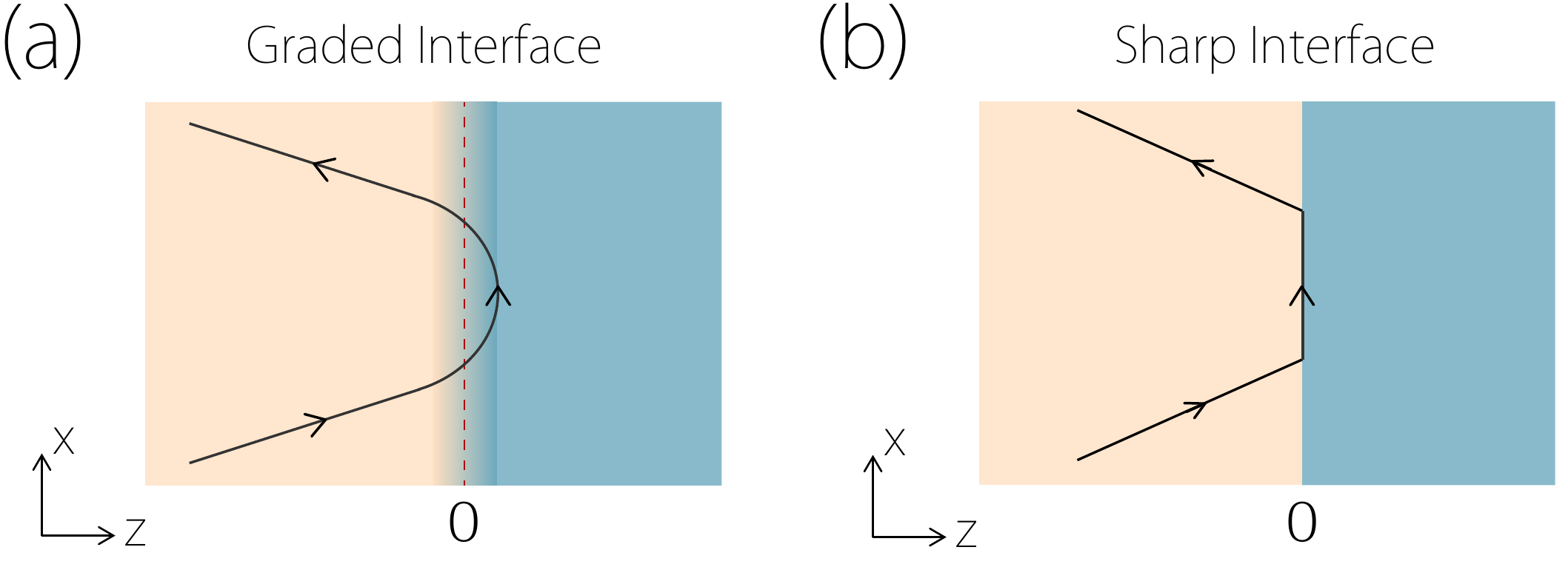}\caption{Schematic figure showing the scattering at (a) a graded interface and (b) a sharp interface.\label{fig3}}
\end{figure}

Before proceeding, we have a few remarks regarding the comparison between the longitudinal shift and the transverse shift. First, the longitudinal shift allows a quite intuitive understanding. Consider a graded interface, where the left medium $H_L$ is smoothly interpolated to the right medium $H_R$. Assuming the right medium does not support a propagating mode for the particle (e.g., it may have a spectral gap at the given particle energy), then the beam will be adiabatically reflected back, and its trajectory is schematically illustrated in Fig.~\ref{fig3}(a). The longitudinal shift can be regarded as due to the bending of the trajectory in the graded interface region. When the width of the graded interface region approaches zero, the interface becomes a sharp one. Then the lateral propagation along the interface is enabled through the evanescent modes at the interface which decays into the right medium, as illustrated in Fig.~\ref{fig3}(b). In comparison, the transverse shift does not have such a simple classical picture, hence appears to be more nontrivial.

Second, due to the above discussion, the longitudinal shift is not well defined for a graded interface, because its value depends on the location taken in the graded region [see Fig.~\ref{fig3}(a)]. The longitudinal shift is only well defined for a sharp interface. In salient contrast, the transverse shift is well defined for both cases, since it is the difference between the incident plane and the scattered plane.

Third, since the longitudinal shift is within the plane of incidence, it can be studied by simply taking a 2D subsystem (without considering the $y$ dimension in Fig.~\ref{fig2}). In contrast, the study of the transverse shift must require a 3D system, i.e., it is a genuine 3D phenomenon.

The first and the third points above might offer a possible explanation for why the transverse shift was revealed much later than the longitudinal shift in electronic systems.

\section{Approaches}

In previous works on the anomalous shifts, three different approaches have been adopted. Each approach has its own applicability, advantage, and limitation. In the following, we review these three approaches.

\subsection{Quantum scattering approach}

The standard and the most general approach is the scattering approach. This is also the traditional approach adopted for studying the anomalous shifts in geometric optics~\cite{BliokhReviw}.

In this approach, one directly solves the scattered particle beam from the incident beam $\bm\Psi^I$. The calculation is facilitated by expanding the beam wave function using the eigenmodes of the system, i.e., the scattering basis states $\psi$. Explicitly, one can write
\begin{equation}\label{Psii}
\bm\Psi^I=\int d\bm k' w(\bm k'-\bm k)\psi^I_{\bm k'}(\bm r),
\end{equation}
where the incident basis state $\psi^I$ (the partial wave) is labeled by its wave vector $\bm k'$. Since we require the beam to be laterally confined, due to the uncertainty principle, it must consist of a spread of basis states, which is described by the beam profile $w$. To have a well defined trajectory, one also needs the profile $w$ to be peaked around an average wave vector $\bm k$. The specific form of $w$ does not affect the final result of the shifts. In calculations, one usually chooses $w$ to have a Gaussian form:
\begin{equation}
w(\bm{q})=\prod_{i}w_{i}(q_{i}),
\end{equation}
where
\begin{equation}
  w_{i}(q_{i})=(\sqrt{2\pi}W_{i})^{-1}e^{-q_{i}^{2}/(2W_{i}^{2})},
\end{equation}
and $W_i$ is the width for the $i$-th component.

When the incident beam hits the interface, it will be scattered. For a concrete discussion, let's assume that there are a single channel for reflection and a single channel for transmission. (The generalization to cases with multiple channels is straightforward.) Then there will be a reflected beam $\bm \Psi^R$ and a transmitted beam $\bm \Psi^T$.

How do we find the reflected beam $\bm \Psi^R$? By using the expansion in Eq.~(\ref{Psii}), we only need to know how each partial wave $\psi^I_{\bm k'}$ is reflected at the interface, which presumably is already known when we obtain the scattering basis states at the first place. Recall that a scattering basis state for the system takes the form of
\begin{eqnarray}
\psi_{\bm k} & = & \begin{cases}
\psi^I_{\bm k}+r(\bm k)\psi^R_{\bm k}, & z<0,\\
t(\bm k)\psi^T_{\bm k}, & z>0.
\end{cases}
\end{eqnarray}
This means that the incident partial wave $\psi^I$ is reflected (transmitted) as $\psi^R$ ($\psi^T$) with an amplitude $r$ ($t$). Consequently, the reflected beam can be obtained as
\begin{equation}
\bm\Psi^R=\int d\bm k' w(\bm k'-\bm k)r(\bm k')\psi^R_{\bm k'}(\bm r).
\end{equation}

The anomalous spatial shift (including both longitudinal and transverse components) can then be obtained by comparing the center positions of $\bm \Psi^R$ and $\bm \Psi^I$ at the interface. For example, if taking $H_L=\hbar^2k^2/(2m)$ to be the simple 3D isotropic electron gas model, for the configuration in Fig.~\ref{fig2}, by expanding the phase of the amplitude $r$ to the first order around $k_y$, one can find that
\begin{equation}
  \bm \Psi^R\propto e^{-W_y^2\big[y+\frac{\partial}{\partial k_y'}\arg(r)\big|_{\bm k_\|}\big]^2/2}.
\end{equation}
Compared with the incident beam $\bm \Psi^I\propto e^{-W_y^2 y^2/2}$, one finds that the reflected beam acquires a transverse shift
\begin{equation}\label{deltay}
\delta y^R=-\frac{\partial}{\partial k_y'}\arg(r)\Big|_{\bm k_\|},
\end{equation}
in the $y$ direction. Similarly, the longitudinal shift can be obtained as $\delta x^R=-\frac{\partial}{\partial k_x'}\arg(r)\big|_{\bm k_\|}$ for this simple model. The expressions for the shifts [like Eq.~(\ref{deltay})] depend on the model, and can become more complicated when the quantum state has some internal spin/pseudospin degree of freedom~\cite{BeenakkerGH2009}. Following similar analysis, one can also find the shifts for the transmitted beam $\bm \Psi^T$.

This approach is based on the analysis of the very fundamental quantum scattering problem. It is quite general. Unlike the semiclassical approach to be reviewed in Sec.~\ref{Semi} which requires the scattering potential to be slowly varying over the particle wavelength~\cite{YangPRL2015}, the quantum scattering approach here does not suffer from this constraint. There is no semiclassical approximation involved. Particularly, it applies for sharp interfaces and for cases when the particle wavelength is relatively large (like for doped semiconductors or semimetals \cite{BeenakkerGH2009,JiangPRL2015,YangPRL2015}).

In practice, it is easier to deal with sharp interfaces, when using the scattering approach, because the mode matching at the interface needed to obtain the scattering amplitudes can be done straightforwardly. For graded interfaces, one has to resort to techniques such as the transfer matrix method to carry out the calculation.

Finally, we mention that the expression in Eq.~(\ref{deltay}) is quite suggestive. It shows that the shift is connected to the variation of the phase angle of the scattering amplitude versus the wave vector. In other words, a finite spatial shift can result from the different phase shifts for the different partial waves in scattering. In addition, when the waves have spin/pseudospin degree of freedom, there could be additional contributions from such internal degree of freedom.

\subsection{Semiclassical approach}\label{Semi}

The second approach is through the application of the semiclassical theory. The semiclassical theory aims to describe the dynamics of a quantum particle using a set of equations of motion analogous to the Hamilton equations in classical mechanics~\cite{XiaoRMP2010}.

One may ask: the uncertainty principle says that you cannot have both position and momentum well defined at the same time, then how can you write down equations to describe their dynamics?
Well, indeed, that is true. When the momentum is precisely defined, like for a plane wave, the position is completely undetermined, and vice versa. To obtain a semiclassical description, we need to make a compromise for both variables, i.e., we let each variable carry certain ``acceptable" uncertainty, such that together they could satisfy the fundamental limit posed by the uncertainty principle. This means that we are studying the dynamics of a particle wave packet $\Psi(\bm r_c,\bm k_c)$. By ``acceptable", the wave packet spread is required to be sufficiently narrow in both position and momentum spaces, such that its center $(\bm r_c,\bm k_c)$ in phase space can be defined.

The validity of the semiclassical description requires the external perturbations to be smooth and slowly varying in space, such that the wave packet can be viewed as a point particle. A guideline is that the length scale for the perturbation must be much greater than the particle wavelength. Thus, regarding our current problem, this means that the semiclassical approach can only apply for graded interfaces, but not for sharp interfaces.

There exists a systematic way to derive the semiclassical equations of motion for $(\bm r_c,\bm k_c)$ from the Schr\"{o}dinger equation for the system. Sundaram and Niu~\cite{SundaramPRB1999} showed that the equations for a Bloch wave packet in a single band take the general form of (setting $\hbar=1$)
\begin{equation}
\dot{\bm r}_c=\frac{\partial \mathcal{E}}{\partial \bm k_c}-(\Omega_{\bm k\bm r}\cdot\dot{\bm r}_c+\Omega_{\bm k\bm k}\cdot\dot{\bm k}_c)-\Omega_{\bm kt},
\end{equation}
\begin{equation}
\dot{\bm k}_c=-\frac{\partial \mathcal{E}}{\partial \bm r_c}+(\Omega_{\bm r\bm r}\cdot\dot{\bm r}_c+\Omega_{\bm r\bm k}\cdot\dot{\bm k}_c)+\Omega_{\bm rt}.
\end{equation}
Here, $\mathcal{E}(\bm r_c,\bm k_c)$ is the energy of the wave packet, and the $\Omega$'s are the various Berry curvatures defined in terms of the gauge potentials $\mathcal{A}$ known as Berry connections. For example, $\mathcal{A}_{q_i}=i\langle u|\partial_{q_i}u\rangle$ ($\bm q=\bm r,\bm k$), where $|u\rangle$ is the periodic part of the Bloch state. Note that the $\bm r_c$ dependence of $|u\rangle$ comes from the dependence of the Hamiltonian on certain spatially varying parameters.
For simple notations, here and hereafter, we drop the subscript $c$ from $\bm r_c$ and $\bm k_c$ whenever appropriate. Then
the phase space Berry curvatures are defined as $\Omega_{k_i r_j}=\partial_{k_i}\mathcal{A}_{r_j}-\partial_{r_j}\mathcal{A}_{k_i}$ ($\Omega_{\bm k\bm k}$, $\Omega_{\bm r\bm r}$, and $\Omega_{\bm r\bm k}$ are similarly defined).  $\Omega_{r_i t}=\partial_{r_i}\mathcal{A}_{t}-\partial_{t}\mathcal{A}_{r_i}$ arises due to certain time dependent parameters in the Hamiltonian, which may lead to pumping effects~\cite{Yang2010}.

For the particular case with weak electric field $\bm E$ and magnetic field $\bm B$. The equations of motion reduce to the form derived by Chang and Niu~\cite{ChangPRL1995,ChangPRB1996} (setting $e=1$):
\begin{equation}
\dot{\bm r}=\frac{\partial \mathcal{E}}{\partial \bm k}-\dot{\bm k}\times \bm \Omega,
\end{equation}
\begin{equation}
\dot{\bm k}=-\bm E-\dot{\bm r}\times\bm B.
\end{equation}
Here $\Omega_\ell\equiv\epsilon^{ij\ell}\Omega_{k_i k_j}/2$ is the familiar momentum space Berry curvature. This set of equations have found wide applications. It should be noted that these equations are accurate to first order in external fields. Gao, Yang, and Niu~\cite{GaoPRL2014,Gao2015} developed a theory with second order accuracy, which can be applied to study nonlinear transport phenomena, such as magneto-transport \cite{Gao2017}.

To apply the semiclassical approach to study the shifts, one can directly solve the trajectory from the evolution of the equations of motion. The advantages of the approach are: (i) The physical picture is very intuitive; (ii) the shifts can be explicitly connected to Berry curvatures, which are intriguing geometric quantities.

The limitations of the approach are also obvious. First, as we have mentioned, it cannot treat sharp interfaces. It follows that this is not a good approach for studying the longitudinal shift (which is only well defined for sharp interfaces). Second, when the wave packet is formed from multiple entangled bands, the description would typically require a non-Abelian multiband treatment~\cite{Culcer2005}, which is more complicated than the equations shown above.

\subsection{Symmetry argument}

In the study of the optical Imbert-Fedorov shift, it was argued that the shift must exist as a result of SOC and total angular momentum conservation \cite{OnodaPRL2004}. Here, the spin is tied to the helicity of the light, and the SOC is inherent in the Maxwell equations~\cite{Bliokh2015}. The total angular momentum conservation is due to the rotational symmetry in the direction normal to the interface.

This approach can also be applied for electronic systems. In a crystalline solid, due to the presence of lattice, we do not have any continuous rotational symmetry. Nevertheless, the effective models which describe the low-energy electrons may acquire an emergent rotational symmetry. For example, the isotropic electron gas model, which may describe the electrons at the conduction band edge for some semiconductors, enjoys a full rotational symmetry along any axis.

Assuming the model of our system possesses such a rotational symmetry along $z$ (as for the setup in Fig.~\ref{fig2}), this means that the total angular momentum, represented by the operator $\hat{\mathcal{J}}_z$, is a conserved quantity:
\begin{equation}
  [\hat{\mathcal{H}},\hat{\mathcal{J}}_z]=0,
\end{equation}
where $\hat{\mathcal{H}}$ is the Hamiltonian for the whole system. Here, we add hats for the symbols to stress that these are operators. Generally, $\hat{\mathcal{J}}_z$ has the form of
\begin{equation}\label{TAM}
\hat{\mathcal{J}}_z=(\hat{\bm r}\times \hat{\bm k})_z+\hat{S},
\end{equation}
where the first term is the orbital angular momentum, and the second term includes any additional contribution which may come from the internal degree of freedom.

Now, considering a wave packet $\Psi^I$ reflected by the interface into $\Psi^R$, the conservation of $\hat{\mathcal{J}}_z$ means that
\begin{equation}
\langle \Psi^I|\hat{\mathcal{J}}_z|\Psi^I\rangle = \langle \Psi^R|\hat{\mathcal{J}}_z|\Psi^R\rangle.
\end{equation}
For the configuration in Fig.~\ref{fig2}, this equation leads to a transverse shift in reflection:
\begin{equation}\label{sym1}
\delta y^R=\frac{1}{k_x}(\langle \Psi^R|\hat{S}|\Psi^R\rangle-\langle \Psi^I|\hat{S}|\Psi^I\rangle).
\end{equation}

The result in Eq.~(\ref{sym1}) shows that the transverse shift is nonzero, when the internal state $\langle \hat S\rangle$, corresponding to spin or some kind of pseudospin, changes in scattering. This change in the internal state generally results from a coupling between spin (pseudospin) and orbital motion. Thus, intuitively, when the angular momentum associated with spin (pseudospin) changes in scattering, the transverse shift must arise so that the change in the orbital angular momentum can compensate to ensure the conservation of the total angular momentum.

When there are multiple scattering channels, the conservation relation above applies for each channel separately. This is required when the particle is regarded as a quantized object, such that each particle is scattering into one of the channels with certain probability.

Certainly, the applicability of this symmetry argument approach requires the presence of the rotational symmetry, which depends on the system. When the symmetry does exist, this approach will be very powerful. From the discussion above, one can see that the result only depends on the asymptotic incident and outgoing wave packet states away from the interface (which are determined by the bulk properties), but not on the details of the interface nor on the detailed interaction between the particle and the interface. It also makes no assumption on the particle wavelength as well as the width of the interface region.

\section{Transverse Shift in Normal Scattering}

Before 2015, most studies on interface scattering, especially on the anomalous shifts in interface scattering, were done for the 2D electronic systems, possibly because such systems can be well fabricated and well controlled, have high mobilities, and are simple enough yet allow nontrivial effects to happen. Nevertheless, regarding the shifts, only the longitudinal shifts can occur for 2D systems, but not the transverse shifts. For a review on these works, please see Ref.~\cite{ChenReview2013}.

The possibility of transverse shifts in electronic interface scattering was first proposed by two works in 2015~\cite{JiangPRL2015,YangPRL2015}. The discovery is a natural byproduct of the study on 3D topological materials. As we have mentioned in Sec.~\ref{Semi}, the transverse shift can be related to the nontrivial Berry curvatures, which are often the key features of topological materials. Particularly, in so-called Weyl semimetals \cite{Wan2011,ArmitageRMP2018}, the conduction and valence bands cross at twofold degenerate Weyl points, which behave as monopole charges for the Berry curvature fields. Hence, the transverse shift has been first revealed for such Weyl electrons in interface scattering.

In a Weyl semimetal, an electron near a Weyl point (at $K_{0}$) may
be described by the effective model
\begin{equation}
H_{0}=-i\chi\sum_{i=x,y,z}v_{i}\sigma_{i}\partial_{i},\label{Weyl}
\end{equation}
where $\chi=\pm1$ is the chirality (also the monopole charge) of the Weyl point, $v_{i}$'s
are the Fermi velocities, $\sigma_{i}$'s are the Pauli matrices corresponding
to a spin or pseudospin degree of freedom. Let us assume that the Weyl points are sufficiently separated in $k$-space, such that the (intervalley) scattering between the different points can be neglected. It is noted that Weyl model in Eq.~(\ref{Weyl}) represents a model with the strongest SOC, because the entire Hamiltonian is of an SOC term~\cite{YangSpin2016}.

In Refs.~\cite{JiangPRL2015, YangPRL2015}, the authors considered one simplest type of interface---the interface caused by an electrostatic potential step $V(z)$. For a sharp interface, one may write $V(z)=V_0\Theta(z)$ for the potential step, where $\Theta(z)$ is the Heaviside step function. According to the setup in Fig.~\ref{fig2}, the model of the system is given by
\begin{equation}
  \mathcal{H}=H_0+V(z).
\end{equation}

To study the transverse shift, Jiang \emph{et al.}~\cite{JiangPRL2015} adopted the quantum scattering approach, while Yang \emph{et al.}~\cite{YangPRL2015} used the symmetry argument. The different approaches reached the same result.


\begin{figure}[t]
\includegraphics[width=8cm]{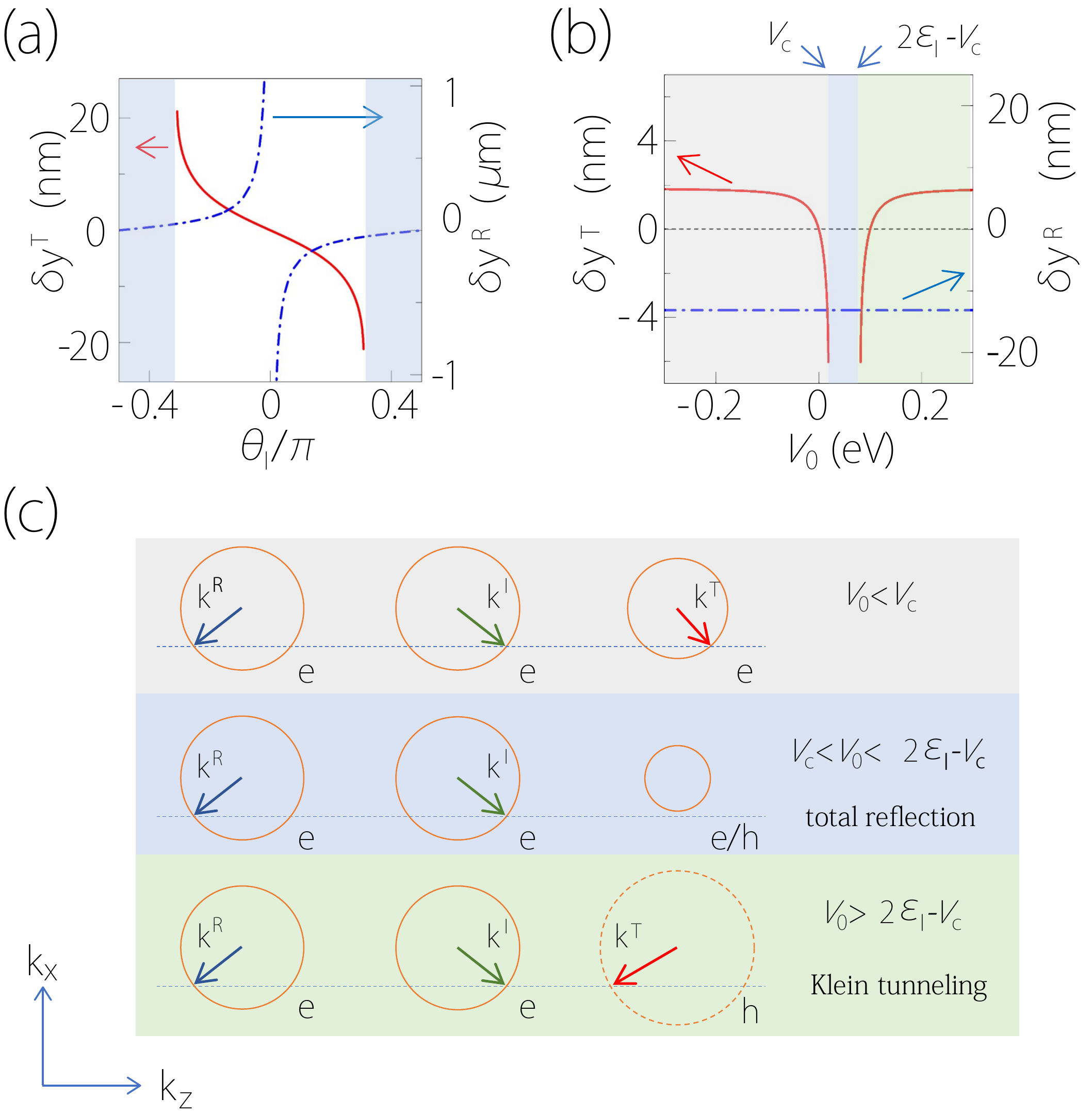}\caption{(a,b) $\delta y^{T,R}$
plotted as functions of the incident angle $\theta_{I}$ and the potential
step height $V_{0}$, respectively. The shaded regions mark the range
of total reflection. (c) Schematic figure showing the Fermi surfaces
and momenta ($k_{y}=0$) of the reflected and/or transmitted wave
packets in normal, total reflection, and Klein tunneling regimes.
The behaviors in (a,b) are for left-handed Weyl electrons, whereas
the behaviors for right-handed ones (not shown) are the opposite;
the features in (c) do not distinguish chirality. Figure adapted with permission from Ref.~\cite{YangPRL2015}. \label{fig4}}
\end{figure}

For example, consider the symmetry argument, which applies when $v_x=v_y$ in model (\ref{Weyl}). For this model, the total angular momentum operator is given by
\begin{equation}
\hat{\mathcal{J}}_z=(\hat{\bm r}\times \hat{\bm k})_z+\frac{\chi}{2}\sigma_z,
\end{equation}
such that its average over the wave packet state is given by
\begin{equation}
\bm J_z=(\bm r\times \bm k)_z+\frac{\chi}{2}(\bm n)_z,
\end{equation}
where $\bm n$ is the unit vector $(v_x k_x, v_y k_y, v_zk_z)/\mathcal{E}_{\bm k}$, and $\mathcal{E}_{\bm k}=\eta\sqrt{v_x^2 k_x^2+v_y^2 k_y^2+v_z^2 k_z^2}$ are the energy dispersions of the electron and hole ($\eta=\pm$) bands.

As sketched in Fig. \ref{fig2} and Fig.~\ref{fig4}(c), consider an incident
electron that has an energy ${\cal E}_{I}>0$ and an incident angle
$\theta_{I}=\arctan(k_{x}^{I}/k_{z}^{I})$ in the $x$-$z$ plane.
All energy scales are assumed to be small compared to the bandwidth,
such that the velocities change little across the step. For $V_{0}<{\cal E}_{I}$,
the transmitted particle ($T$) is also a Weyl electron. The energy
and momentum conservations require that ${\cal E}_{I}=V_{0}+{\cal E}_{T}$
and $k_{x}^{I}=k_{x}^{T}$. The conservation of $J_{z}$ further
determines the transverse shift. For the transmission process, one obtains
\begin{equation}
\delta y^{T}=\chi\frac{(n_{z}^{T}-n_{z}^{I})}{2k_{x}^{T}}=\chi\frac{v_{z}}{2}\left(\frac{\cot\theta_{T}}{{\cal E}_{I}-V_{0}}-\frac{\cot\theta_{I}}{{\cal E}_{I}}\right),\label{eq:Tshift}
\end{equation}
where $\theta_{T}=\arctan(k_{x}^{T}/k_{z}^{T})$ is the refraction
angle. When $V_{0}>V_{c}\equiv{\cal E}_{I}-v_{x}|k_{x}^{I}|$ becomes
imaginary and a total reflection occurs.

For $V_{0}>{\cal E}_{I}$, the potential step forms a $p$-$n$ junction
and Klein tunneling of the Weyl electron may occur. Evidently, when
${\cal E}_{I}<V_{0}<2{\cal E}_{I}-V_{c}$, $k_{z}^{T}$ is imaginary
and the total reflection occurs. Yet, when $V_{0}>2{\cal E}_{I}-V_{c}$,
Klein tunneling is possible and the minus sign in $k_{z}^{T}$ indicates
a positive group velocity of the outgoing hole. Interestingly, a negative refraction occurs in this case, \emph{i.e.}, $\theta_{I}\theta_{T}<0$. Meanwhile, the transverse
shift takes the same form as in Eq.~(\ref{eq:Tshift}).

The transverse shift in reflection ($R$) can be obtained in a similar way, given by
\begin{equation}
\delta y^{R}=\chi\frac{(n_{z}^{R}-n_{z}^{I})}{2k_{x}^{R}}=-\frac{\chi v_{z}\cot\theta_{I}}{{\cal E}_{I}},\label{eq:Rshift}
\end{equation}
where $k_{x}^{R}=k_{x}^{I}$ and $k_{z}^{R}=-k_{z}^{I}$.

The dependence of these shifts on the incident angle and the potential step is shown in Fig.~\ref{fig4}(a) and \ref{fig4}(b). One notes the following points. First, the shifts are odd functions of the incident angle. Second, when the symmetry argument holds, the transverse shifts would have universal behaviors independent of the interface details. Particularly, the results above apply for both sharp and graded interfaces. Third, the sign of the shifts depends on the chirality $\chi$. Hence, the Weyl electrons with different chiralities should shift in opposite directions. This leads to the proposition of the chirality Hall effect in Ref.~\cite{YangPRL2015}.

It is also noted that the shift in reflection diverges when $\theta_I$ approaches perpendicular incidence. Physically, the shift cannot diverge. There are two factors that regulate this diverging behavior. (i) The probability of reflection is completely suppressed at perpendicular incidence due to the reversed spin direction, so the seemingly diverging shift at perpendicular incidence cannot manifest in measurement. (ii) Due to the uncertainty principle, a confined beam must have a finite spread in
the wave vector (and hence the incident angle) distribution for the partial waves. When approaching perpendicular incidence, the diverging behavior indicates that the different partial waves would scatter in drastically different ways, such that the scattered beam would no longer be confined and the shift would become ill-defined. Consequently, although the shift should get enhanced with decreasing incident angle, the diverging behavior at perpendicular incidence would not occur in reality.

The results above can be exactly reproduced by the quantum scattering approach. As mentioned, the scattering approach is more general, and it applies also for cases without the rotational symmetry. For example, if $v_x\neq v_y$ in the current model, one finds from the scattering approach that \cite{JiangPRL2015}
\begin{eqnarray}\label{yR}
\delta y^{R} & = & -\chi\frac{v_{x}v_{z}}{v_{y}}\cdot\frac{\cot\theta_{I}}{{\cal E}_{I}}
\end{eqnarray}
for reflection, which recovers Eq.~(\ref{eq:Rshift}) when $v_x\rightarrow v_y$. In fact, for the simple Weyl model studied here, the result
in Eq.~(\ref{yR}) can also be obtained from the symmetry argument after making a scaling transformation on the coordinates: $(x',y')=(x\sqrt{v_y/v_x},y\sqrt{v_x/v_y})$.

Finally, for graded interfaces, the transverse shift can also be obtained from the semiclassical approach. As derived in Ref.~\cite{YangPRL2015}, assuming the potential $V$ and the Fermi velocities $v$'s are slowly varying spatially compared to the Fermi wavelength,
the semiclassical equations of motion for the Weyl wave packet center $(\bm r,\bm k)$ take the following form
\begin{equation}
\dot{\bm r}=\frac{\partial \mathcal{E}}{\partial \bm k}-\Omega_{\bm k\bm r}\cdot\dot{\bm r}-\dot{\bm k}\times\bm \Omega,
\end{equation}
\begin{equation}
\dot{\bm k}=-\frac{\partial \mathcal{E}}{\partial \bm r}-\frac{\partial V}{\partial \bm r}+\Omega_{\bm r\bm k}\cdot\dot{\bm k}.
\end{equation}
For the Weyl model in Eq.~(\ref{Weyl}), the momentum space Berry curvature is given by
\begin{equation}
\bm{\Omega}=-\chi\frac{v_{x}v_{y}v_{z}\bm{k}}{2\mathcal{E}^3},
\end{equation}
for the electron state in the conduction band.

The shift can be determined by integrating these two equations of motion. For example, when the $v$'s are constants and $V$ depends on $z$ only, the anomalous shift in the $x$-$y$ plane is given by
\begin{equation}
\delta\bm\ell^\alpha=-\int_I^\alpha \dot{\bm k}\times\bm \Omega\ dt,
\end{equation}
where $\alpha=T$ for $V_0<V_c\equiv \mathcal{E}_I-\sqrt{v_x^2 k_x^2+v_y^2 k_y^2}$ and $\alpha=R$ for $V_0>V_c$. This equation shows that the shift is closely connected with the Berry curvature. In the presence of rotational symmetry ($v_x=v_y$), one can check that it leads to the same results in Eqs.~(\ref{eq:Tshift}) and (\ref{eq:Rshift}). However, there are two important points to be noted. First, the \emph{semiclassical} trajectory is unique, i.e., transmission if $V_0<V_c$ and reflection if $V_0>V_c$. The transmission and reflection cannot occur simultaneously like in the quantum case. Second, the above equation cannot apply to the Klein tunneling case because the point where $\mathcal{E}=0$ requires a non-Abelian treatment.

On the other hand, the spatial ($z$-)variation of the Fermi velocities alone can also lead to an anomalous shift. From the equations of motion, one finds that \cite{YangPRL2015}
\begin{equation}
  \delta\bm\ell^\alpha=-\int_I^\alpha\frac{1}{1+\Omega_{k_z z}}\Big[\Omega_{\bm k z}\frac{\partial \mathcal{E}}{\partial k_z}+(\bm \Omega\times \hat{z})\frac{\partial \mathcal{E}}{\partial z}\Big]dt.
\end{equation}
One notes that apart from the contribution due to the momentum space Berry curvature $\bm \Omega$, there is an additional contribution entirely due to the phase space Berry curvature $\Omega_{\bm k\bm r}$. The phase space Berry curvature is less well known. The transverse shift is probably the first predicted physical effect induced by $\Omega_{\bm k\bm r}$.

From the semiclassical approach, one can see that the transverse shift should generally exist for materials with nontrivial Berry curvatures. Hence it is not limited to Weyl semimetals. There are many different types of topological semimetals discovered in recent years, which may also give rise to transverse shifts. For example, the transverse shifts in multi-Weyl semimetals have been studied in Ref.~\cite{WangPRB2017}, and the different behaviors for intravalley and intervalley scattering processes have been addressed.

It should be mentioned that a sizable longitudinal shift also exists for the Weyl electron scattering, which has been investigated using the scattering approach for sharp interfaces. Later, Jiang \emph{et al.}~\cite{JiangPRB2016} proposed that these anomalous shifts (longitudinal and transverse) lead to an anomalous scattering probability for a Weyl wave packet scattered at defect potentials, which enhances the ratio between the transport lifetime and the quantum lifetime. Intuitively, the anomalous shift helps the Weyl electron to circumvent the scatterer, effectively reducing the strength of disorder scattering. This was suggested as a possible explanation for the high mobility observed for Weyl semimetal materials.

\section{Transverse Shift in Andreev Reflection}

\begin{figure}[t]
\includegraphics[width=8cm]{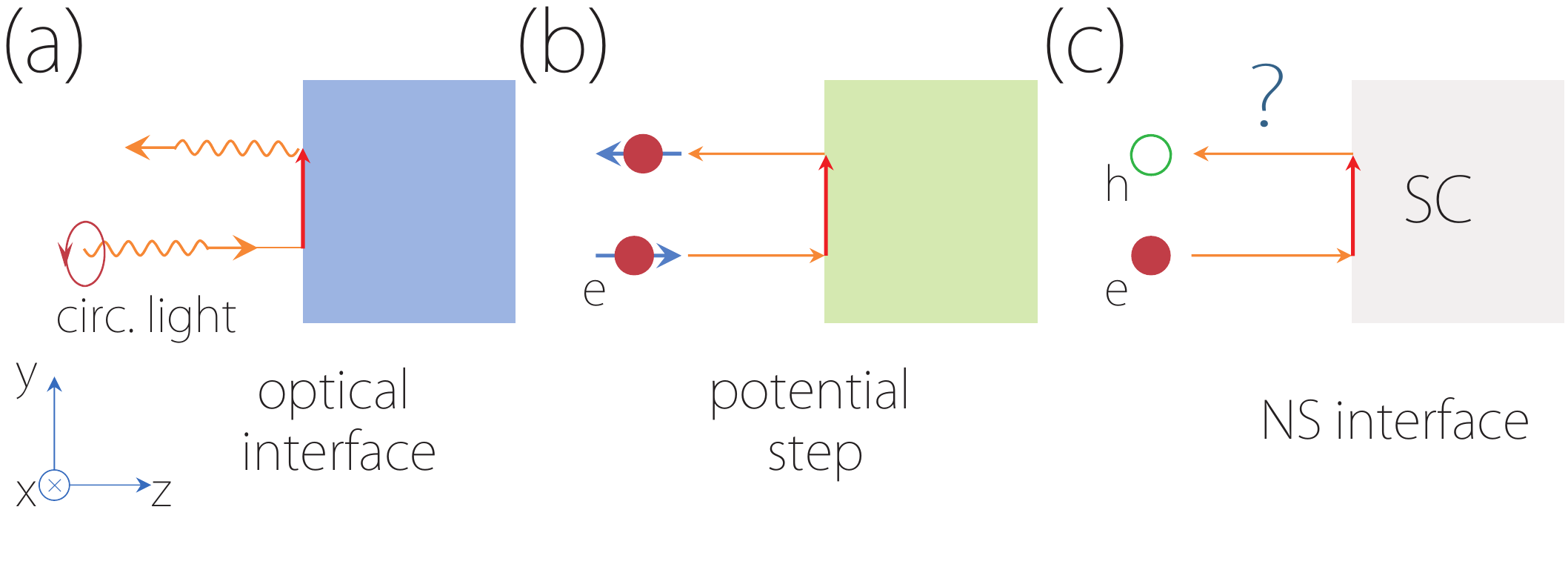}\caption{Three kinds of reflection processes. (a) A circularly polarized light beam undergoes a transverse shift (Imbert-Fedorov effect) when reflected at an optical interface. (b) Electron with strong spin-orbit coupling, like in Weyl semimetals, acquires a transverse shift when reflected from a potential barrier. (c) An incident electron is reflected as a hole in Andreev reflection from a normal-metal/superconductor (NS) interface. \label{fig7a}}
\end{figure}

In the previous examples, the scattered particle and the incident particle are of the same identity: a photon is scattered as another photon, and an electron is scattered as another electron. However, there is an intriguing scattering process happening at the interface between a normal-metal (N) and a superconductor (S), in which the particle identity is changed [see Fig.~\ref{fig7a}(c)]. This is the famous Andreev reflection~\cite{Andreev1964,Gennes1966}.

In Andreev reflection, an incoming electron from the N side at excitation energy $\varepsilon$
above the Fermi level $E_{F}$ is reflected back as a hole
with energy $\varepsilon$ below $E_{F}$. The
process conserves energy and momentum but not charge: the missing
charge of $(-2e)$ is absorbed as a Cooper pair into
the superconductor. For excitation energies below the superconducting
gap, electrons cannot penetrate into the superconductor, and Andreev
reflection becomes the dominating mechanism for transport through the NS interface.

\emph{Is there any transverse shift in Andreev reflection?} At first sight, this seems unlikely, because the incident and the scattered particles are of distinct identities, even their electric charges are opposite. However, it should be noted that the two particles are not independent. They do correlate coherently through the superconductor on the other side of the interface.

\subsection{Junction with Conventional Superconductor}\label{NSC}

The question above was first addressed by Liu, Yu, and Yang in 2017~\cite{LiuSUTDPRB2017}, with an affirmative answer. They investigated NS junctions consisting of a normal metal with strong SOC and a conventional $s$-wave superconductor. The conclusion is that a finite transverse shift generally exists in Andreev reflection, which is connected with the SOC in the normal metal.

Mathematically, the model is not much different from that for the normal-state junctions, except that the scattering here is governed by the Bogoliubov-de Gennes (BdG) equation~\cite{Gennes1966} instead of the Schr\"{o}dinger equation. For the N side ($z<0$), one has
\begin{eqnarray}
\mathcal{H}_{L} & = & \left[\begin{array}{cc}
H_{L}-E_{F} & 0\\
0 & E_{F}-\mathcal{T}^{-1}H_{L}\mathcal{T}
\end{array}\right];
\end{eqnarray}
whereas for the S side ($z>0$),
\begin{eqnarray}
\mathcal{H}_{R} & = & \left[\begin{array}{cc}
H_{R}-E_{F} & \Delta\\
\Delta^{*} & E_{F}-\mathcal{T}^{-1}H_{R}\mathcal{T}
\end{array}\right].
\end{eqnarray}
Here, $H_L$ and $H_R$ are the Hamiltonian for the normal states of the two sides, $\mathcal{T}$ is the time reversal operator, and $\Delta$ is the superconducting pair potential which couples electron and hole excitations. The BdG Hamiltonian for the whole system may be written as
\begin{equation}
  \mathcal{H}_\text{BdG}=\mathcal{H}_{L}\Theta(-z)+\mathcal{H}_{R}\Theta(z),
\end{equation}
where $\Theta$ is the Heaviside step function. The scattering states are solutions of the BdG equation
\begin{equation}
  \mathcal{H}_\text{BdG}\psi=\varepsilon\psi,
\end{equation}
where the wave function $\psi\equiv(u,v)^T$ is a multicomponent spinor with $u$ ($v$) standing for the electron (hole) state.

The above treatment assumes a sharp interface. The step function model for the pair potential has been widely used in literature~\cite{Blonder1982,Jong1995,Kashiwaya2000}, and it has been shown to be a good approximation
to the full self-consistent solution of the BdG equation for such junction structures~\cite{Plehn1991,Hara1993,Plehn1994}. Particularly, it is accurate when there is
large Fermi momentum mismatch across the interfaces (which effectively reduces the coupling between the layers). On the S side, the mean-field requirement for superconductivity is that the Fermi wavelength in S should be much smaller than the coherence length. It should be noted that the
Fermi wavelength in N is not constrained to be small. Particularly, when N is of a doped semiconductor or semimetal, one may have the Fermi energy on the N side comparable to $\Delta_0\equiv|\Delta|$. The possible existence of an interface barrier can also be described in the model by adding a term $h\delta(z)\tau_z$, where $h$ represents the barrier strength and $\tau$'s are the Pauli matrices for the Nambu space. This barrier mainly affects the scattering probabilities~\cite{Blonder1982}.

In Ref.~\cite{LiuSUTDPRB2017}, the authors demonstrated the transverse shifts using two concrete models. The first is based on the Weyl semimetal model, by letting $H_L=H_0$ and $H_R=H_0-U_0$, where $H_0$ is the Weyl model in Eq.~(\ref{Weyl}) and $U_0$ is some constant potential offset. Here, $U_0$ is needed to fulfill the mean-field requirement for superconductivity on the S side (such that $E_F+U_0\gg \Delta_0$). In the BdG Hamiltonian, an electron state at $\bm k$ is related to a hole state at $-\bm k$. If the $\mathcal{T}$ symmetry is assumed for the system, then the reflected Weyl hole should have the same chirality as the incident Weyl electron.

Via the quantum scattering approach, the transverse shift in Andreev reflection was derived for such Weyl NS junction model, given by
\begin{equation}
\delta y_\text{A}=\frac{\chi}{2}\frac{v_yv_{z}}{v_x}\left(\frac{\cot\theta_{h}}{E_{F}-\varepsilon}-\frac{\cot\theta_{e}}{E_{F}+\varepsilon}\right),\label{sym}
\end{equation}
where $\theta_{e/h}=\arctan(k_x/k_z^{e/h})$ is the incident/reflection angle.

The result can also be reached with the symmetry argument when $v_x=v_y$. In the current case, one finds that the conserved quantity is
\begin{equation}
\hat{\mathcal{J}}_{z}=(\hat{\bm{r}}\times\hat{\bm{k}})_{z}+\frac{\chi}{2}\tau_{0}\otimes\sigma_{z}.
\end{equation}
Recall that $\tau$ and $\sigma$ are for the Nambu and spin spaces, respectively. The conservation leads to the transverse shift
\begin{equation}
\delta y_\text{A}=\frac{\chi}{2k_{x}}(n_{z}^{h}-n_{z}^{e}),
\end{equation}
where
\begin{equation}
  \bm{n}^{e/h}=(v_{x}k_{x},v_{y}k_{y},v_{z}k_{z}^{e/h})/(E_{F}\pm\varepsilon)
\end{equation}
is the spin polarization direction for the electron/hole.

\begin{figure}[t]
\includegraphics[width=7.8cm]{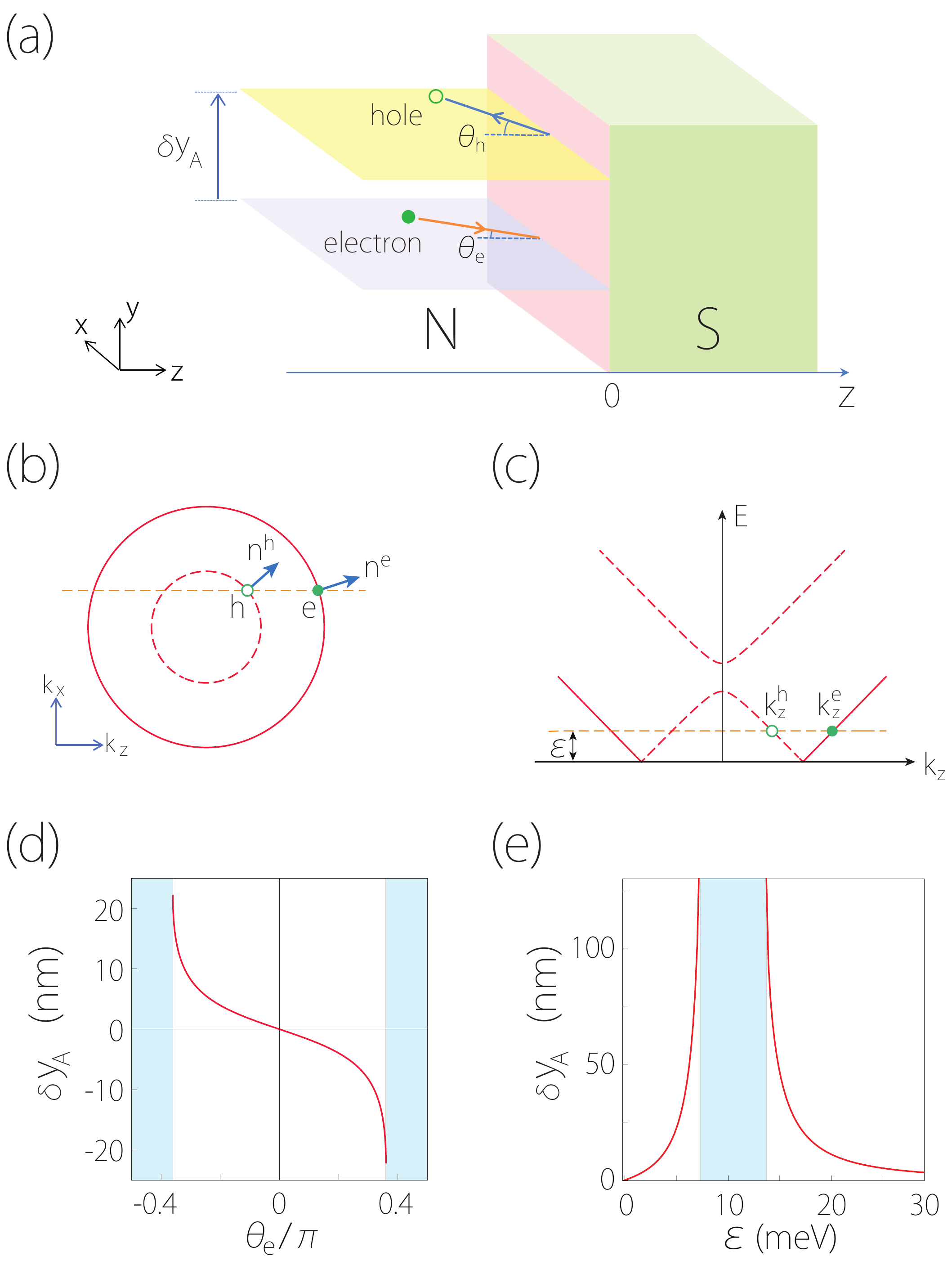}\caption{(a) Schematic figure showing the transverse shift $\delta y_{A}$
for an incident electron wave-packet in the $x$-$z$ plane Andreev
reflected at the NS interface. (b,c) Schematic figure showing (b)
the BdG Fermi surfaces, and (c) spectrum at a finite $k_{x}$ {[}corresponding
to the horizontal dashed line in (b){]}. The solid (hollow) sphere
denotes the incident electron (reflected hole) state, and the arrows
indicate their spin polarization directions. (d,e) Transverse shift versus (d)
incident angle $\theta_{e}$ and (e) excitation energy.
Figure adapted with permission from Ref.~\cite{LiuSUTDPRB2017}.
\label{fig: AR_Weyl_S}}
\end{figure}

Typical behavior of the shift is shown in Fig.~\ref{fig: AR_Weyl_S}(d-e). One observes that $\delta y_\text{A}$ is an odd function
of the incident angle $\theta_{e}$; it vanishes at normal incidence
where $\bm{n}_{e}$ and $\bm{n}_{h}$ are parallel, and reaches maximum
magnitude at a critical angle $\theta_{e}^{c}$,
beyond which $k_{z}^{h}$ becomes imaginary and electrons can no longer
be Andreev reflected. $\delta y_\text{A}$ vanishes when $\varepsilon\ll E_{F}$
or $\varepsilon\gg E_{F}$, because $\bm{n}_{e}$ and $\bm{n}_{h}$
become parallel in both limits; and its seemingly divergent behavior
at $\varepsilon\rightarrow E_{F}$ is reconciled by noting that in
this limit the hole Fermi surface becomes a point, so the reflection
has a vanishingly small probability. In fact, $\varepsilon=E_{F}$
marks the transition point between Andreev retroflection ($\theta_{e}\theta_{h}>0$)
and specular reflection ($\theta_{e}\theta_{h}<0$), as first studied
in graphene~\cite{Beenakker2006}. One finds that $\delta y_\text{A}$ has the same sign in both regimes. Importantly,
the shift is opposite for different chirality, which may generates
a chirality Hall effect also for the Andreev-reflected holes,
similar to that for the normal reflection~\cite{YangPRL2015}.

It should be emphasized that as long as the symmetry argument is valid, the transverse
shift is \emph{independent} of the details of the NS interface and
of the S region for the current setup. This indicates that the Weyl-like model as well as SOC are \emph{not necessary} for
the S region; the effect results from the SOC on the N side.

\begin{figure}[t]
\includegraphics[width=8cm]{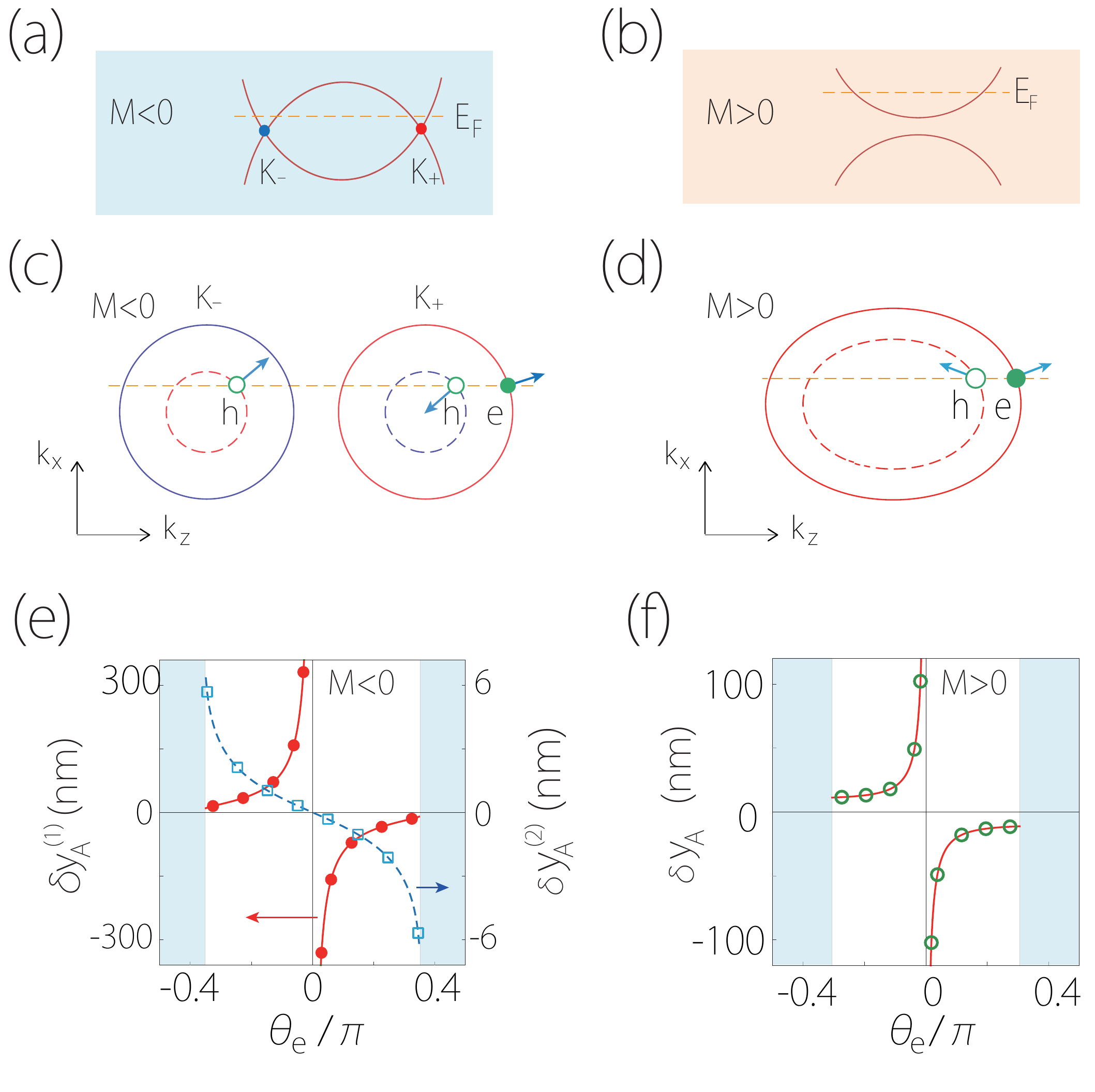}\caption{
(a,b) Two phases of model (\ref{M2}) and their corresponding BdG Fermi
surfaces are schematically shown in (c) and (d), respectively. (e) Shift
$\delta y_{A}^{(1)}$ ($\delta y_{A}^{(2)}$) for intravalley (intervalley)
Andreev reflection versus the incident angle $\theta_{e}$ for $M<0$,
with incident electron from the $K_{+}$ valley. (f) shows the corresponding
result for $M>0$. In (e,f), the data points are from scattering approach,
while the curves are from symmetry argument.
Figure adapted with permission from Ref.~\cite{LiuSUTDPRB2017}.
\label{fig: AR_semiC_S}}
\end{figure}

Another important question is: Does the shift only exist for junctions with Weyl or other topological semimetals? From the symmetry argument, one can see that the answer is negative. Liu, Yu, and Yang ~\cite{LiuSUTDPRB2017} demonstrated this with a concrete example. They considered the following model
\begin{equation}
H_{L}=\frac{1}{2m_L}(-\nabla^{2}+M)\sigma_{z}-iv\sigma_{x}\partial_{x}-iv\sigma_{y}\partial_{y},\label{M2}
\end{equation}
and
\begin{equation}
H_{R}=\Big(-\frac{1}{2m_R}\nabla^{2}-U_{0}-E_{F}\Big)\sigma_{0},
\end{equation}
where $m_{L}$ and $m_R$ are the effective masses for the two sides, $M$ and $v$ are model parameters.  The advantage of $H_L$ is that it
nicely interpolates between two distinct phases determined by
the sign of $M$: for $M<0$, it has a pair of Weyl points on the $k_{z}$ axis at $\pm\sqrt{-M}$
with \emph{opposite} chirality, which simulates a $\mathcal{T}$-broken
Weyl semimetal [see Fig.~\ref{fig: AR_semiC_S}(a)]; for $M>0$, the
two bands are fully separated with a gap [see Fig.~\ref{fig: AR_semiC_S}(b)], and when $E_{F}>M/2m_{L}$, it becomes a doped semiconductor, without any band crossing.
As for the S side, we take it to be the simplest metallic superconductor. In this model, the SOC appears on the N side, but not the S side.

For $M<0$, the transverse shift exists and can be calculated similar to the Weyl model, as shown in Fig.~\ref{fig: AR_semiC_S}(e). One interesting point here is that the shift can happen for both intravalley and intervalley scattering processes, but with distinct dependence on the incident angle  [see Figs.~\ref{fig: AR_semiC_S}(c) and \ref{fig: AR_semiC_S}(e)]. More interestingly, the transverse shift still exists for the doped semiconductor case ($M>0$) [see Figs. \ref{fig: AR_semiC_S}(d) and \ref{fig: AR_semiC_S}(f)]. The value is given by
\begin{equation}
  \delta y_\text{A}=\frac{1}{2k_{x}}(n_{z}^{h}-n_{z}^{e}),
\end{equation}
with
\begin{equation}
n_{z}^{e/h}=\pm[(E_{F}\pm\varepsilon)^{2}-v^{2}k_{\|}^{2}]^{\frac{1}{2}}/(E_{F}\pm\varepsilon).
\end{equation}

These results explicitly demonstrate the following points. (i) The key ingredient
here is the SOC on the N side, however, Weyl or other types of
band crossings are not necessary. (ii) The role of the S side is to
enable the electron-hole conversion. Any conventional superconductor
suffices and it does not require SOC. (iii) Factors such as intervalley
scattering, interfacial barrier, Fermi surface mismatch, and spatial
profile of the pair potential are inessential for the shift. And when
symmetry argument applies, they have no effect on the value of the
shift, although they do affect the probability of the process. In addition, we emphasize again that
the $\sigma$ here can be real spin or any pseudospin. A transverse shift should be induced, as long as the spin state is coupled with
the orbital motion and is changed in the scattering. Particularly, the
results for the Weyl model should directly apply for those spin-orbit-free
Weyl semimetals~\cite{Weng2015,Chen2015}.

\subsection{Junction with Unconventional Superconductor}

For the junctions discussed in Sec.~\ref{NSC}, the key ingredient is the SOC on the N side, which is similar to cases in optics and in normal electron scattering. The shift would vanish if the SOC is negligible. Meanwhile, the superconductor only plays a \emph{passive} role, i.e., a channel for electron-hole conversion.

In a following work, Yu \emph{et al.}~\cite{YuSUTDPRL2018} discovered that a fundamentally new effect can appear for junctions with unconventional superconductors, as illustrated in Fig. \ref{fig7a}(c). There, by ``unconventional", the authors referred to superconductors with unconventional pair potentials.

The key observation is that unconventional pair potentials necessarily have a strong wave-vector dependence~\cite{Sigrist2005}. This generates an \emph{effective} coupling between the orbital motion and the pseudospin of the Nambu (electron-hole) space. Thus, the transverse shift can arise even \emph{in the absence of SOC}. Remarkably, Yu \emph{et al.}~\cite{YuSUTDPRL2018} found that the behavior of the shift is sensitive to the structure of the pair potential
and manifests characteristic features for each pairing type, as summarized in Table~\ref{T1}. Therefore, the effect may provide a powerful new technique capable of probing the structure of unconventional pairings.

{\renewcommand{\arraystretch}{1.2}
{\begin{table}[b]
\newcommand{\tabincell}[2]{\begin{tabular}{@{}#1@{}}#2\end{tabular}}
\centering
\begin{tabular}{ccccccc}
 \hline\hline
   \multicolumn{2}{c}{\multirow{3}{*}{\tabincell{c} {\\Pair \\ potential}}}&
 \multicolumn{2}{c}{Expression}&
  \multirow{3}{*}{\tabincell{c}{\\Period \\ in $\alpha$}} &
  \multirow{3}{*}{\tabincell{c} {\\vanish for  \\ $\varepsilon>|\bm{\Delta}(\alpha)|$}} &
 \multirow{3}{*}{\tabincell{c}{\\No. of  \\ SZ}}
  \\
    \cline{3-4}
    \multicolumn{2}{c}{} & {} & {} &  {} & {}&{}\\
    \multicolumn{2}{c}{} & $\delta\ell_{T}^{e}$ & $\delta\ell_{T}^{h}$ &  {} & {}&{}\\
 \hline
 Chiral & $\Delta_0 e^{i\chi \phi_k}$ & 0 & $\frac{\chi}{k_F\sin \gamma}$ & $\diagup$ &No & 0 \\
 $p_x$ & $\Delta_0 \cos \phi_k$ & \multicolumn{2}{c}{\multirow{4}{*}{$\delta\ell_T^e \approx  \delta\ell_T^h $}} & $\pi$ &\multirow{4}{*}{Yes} & 2 \\
  $p_y$& $\Delta_0 \sin \phi_k$ & \multicolumn{2}{c}{} & $\pi$ & {} & 2 \\
  $d_{x^2-y^2}$& $\Delta_0 \cos 2\phi_k$ & \multicolumn{2}{c}{} & $\pi/2$ & {} & 4 \\
  $d_{xy}$& $\Delta_0 \sin 2\phi_k$ & \multicolumn{2}{c}{} & $\pi/2$ & {} & 4 \\
 \hline\hline
 \end{tabular}
\caption{Features of the transverse shift for typical unconventional pair potentials. ``No. of SZ" stands for the number of suppressed zones when the rotation angle $\alpha$ varies from 0 to $2\pi$. Reproduced from Ref.~\cite{YuSUTDPRL2018}.}\label{T1}
\end{table}}
}

To demonstrate the effect, they took a simplest model, with
\begin{equation}
  H_{L}=-\frac{1}{2m}\nabla^{2},
\end{equation}
and
\begin{equation}
H_{R}=-\frac{1}{2m_{\|}}(\partial_{x}^{2}+\partial_{y}^{2})-\frac{1}{2m_{z}}\partial_{z}^{2}-U_0.
\end{equation}
On the S side, there are two effective mass parameters $m_\|$ and $m_z$. This is for describing the possible anisotropy in the S material.
For certain layered superconductors (like cuprates), the Fermi
surface is highly anisotropic and may take a cylinder-like shape in
the normal state. Such cases can be described by using a lattice model. For unconventional superconductor, the pair potential $\Delta$ in the BdG Hamiltonian would have characteristic wave vector dependence. Often one considers the weak coupling limit, with $E_{F}+U_0\gg |\Delta|,\varepsilon$
in the S region, so that the wave vector for ${\Delta}$'s $k$-dependence
is fixed on the (normal state) Fermi surface of S, and ${\Delta}$
only depends on the direction of the wave vector $\bm{k}$~\cite{Tanaka1995}.

\begin{figure}[t]
\includegraphics[width=8cm]{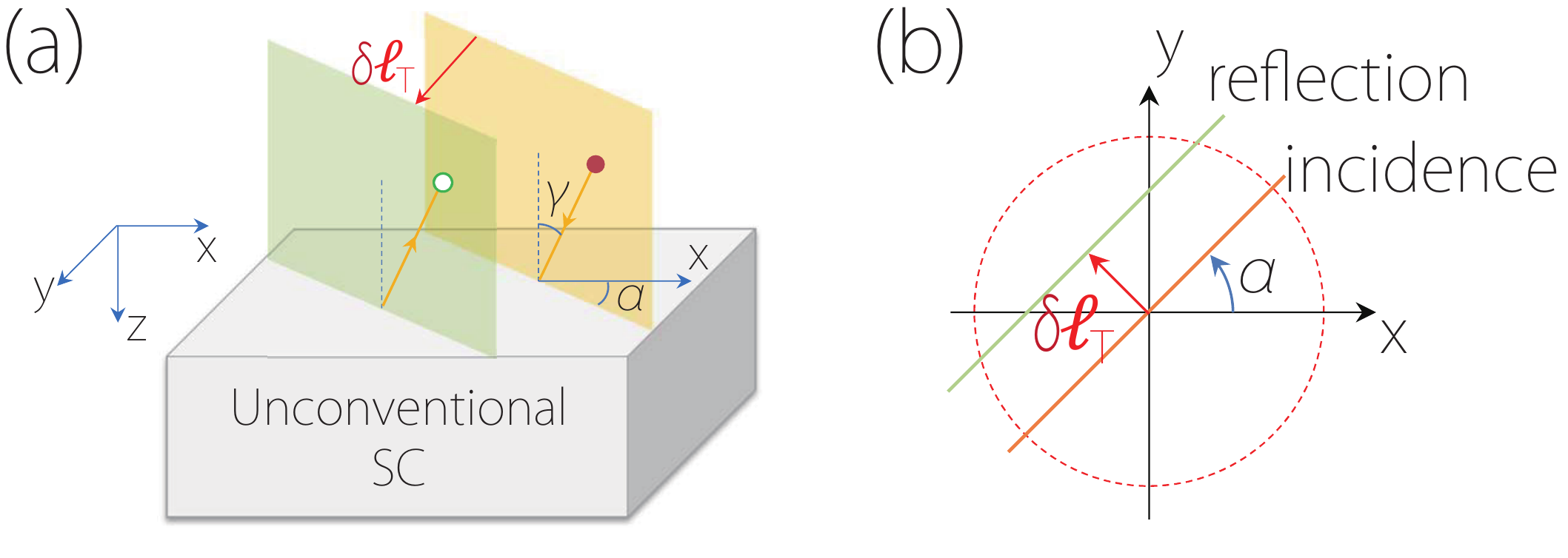}\caption{(a) Schematic of the NS junction set-up. In Andreev reflection, the
reflection plane (green-colored) is shifted by distance $\delta\ell_{T}$
from the incident plane (orange-colored) along its normal direction
($\hat{n}$), due to unconventional pairing in S. (b) Top view of the $x$-$y$ plane in (a). For
certain pairings, there may also be a finite shift for normal-reflected electrons (not shown here). Figure adapted with permission from Ref.~\cite{YuSUTDPRL2018}. \label{fig: ill_junct}}
\end{figure}

The symmetry argument can be applied for the chiral $p$-wave
pairing case, with ${\Delta}=\Delta_{0}e^{i\chi\phi_{k}}$. Here,
$\chi=\pm1$ denotes the chirality of the pairing, and $(\theta_{k},\phi_{k})$ are the
spherical angles of $\bm{k}$. The magnitude $\Delta_{0}$ is assumed to be
independent of $\phi_{k}$ but may still depend on $\theta_k$.
One finds that the quantity
\begin{equation}
\hat{\mathcal{J}}_{z}=(\hat{\bm{r}}\times\hat{\bm{k}})-\frac{1}{2}\chi{\tau}_{z}
\end{equation}
resembles an effective angular momentum operator, and it commutes with the BdG Hamiltonian $\mathcal{H}_\text{BdG}$. For electrons
and holes, the expectation values of the Nambu pseudospin are opposite:
$\langle\hat{\tau}_{z}\rangle_{e/h}=\pm1$. Because the pseudospin
flips in Andreev reflection, the conservation of $J_{z}$ must dictate
a transverse shift $\delta \ell_{T}$ to compensate this change. (Here, the plane of incidence is not assumed to be the $x$-$z$ plane, so the symbol $\delta \ell_T$ instead of $\delta y$ is used to denote the transverse shift. See
Fig.~\ref{fig: ill_junct}.) The result is
\begin{equation}
\delta \ell_T=-\frac{\chi}{2k_{\|}}(\langle{\tau}_{z}\rangle_{h}-\langle{\tau}_{z}\rangle_{e})=\frac{\chi}{k_{\parallel}},\label{eq:chiralp}
\end{equation}
where $k_{\|}=k_{F}\sin\gamma$, $k_{F}$ is the Fermi wave vector
in N, and $\gamma$ is the incident angle.

\begin{figure}[t]
\includegraphics[width=8cm]{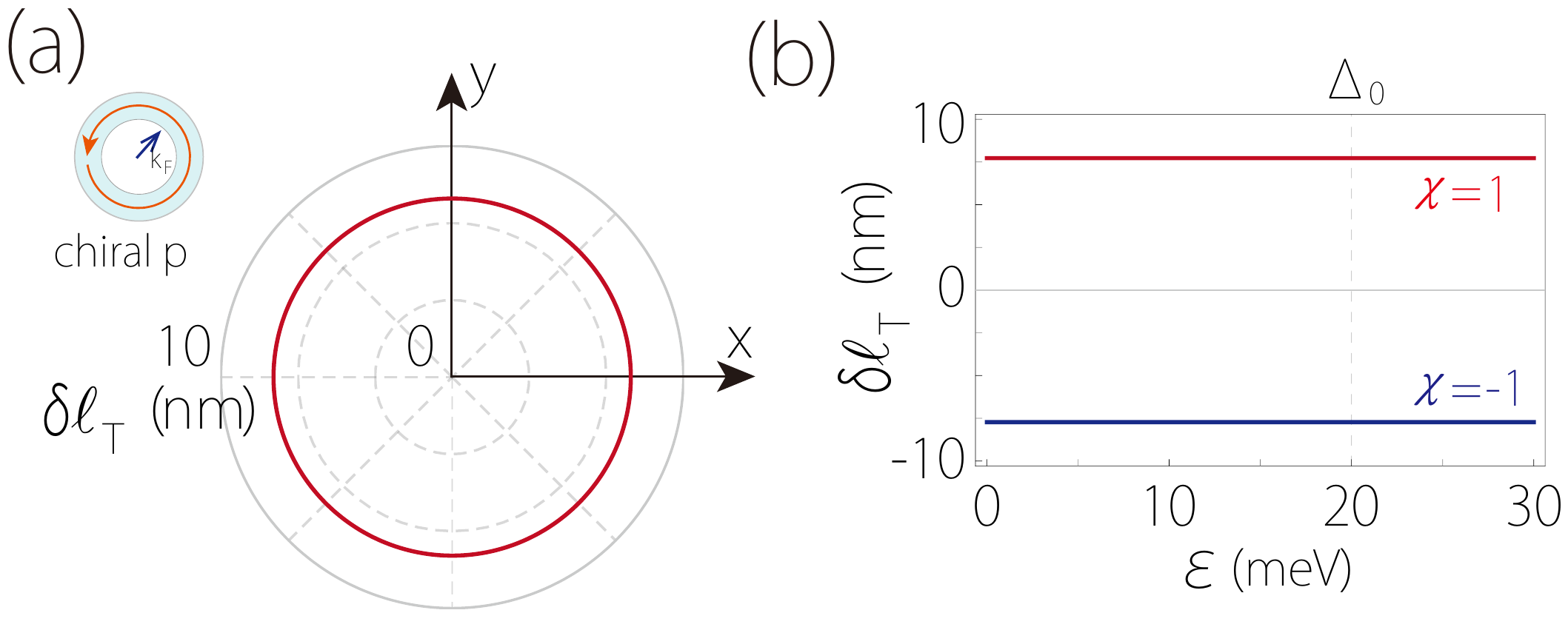}\caption{Transverse shift $\delta\ell_{T}$ for chiral $p$-wave pairing versus
(a) rotation angle $\alpha$ (here $\chi=+1$), and (b) the excitation
energy $\varepsilon$. $\delta\ell_{T}$ is independent of $\varepsilon$
and $\Delta_{0}$, and its sign depends on $\chi$.
Figure adapted with permission from Ref.~\cite{YuSUTDPRL2018}. \label{fig: AR_USC_pwave}}
\end{figure}

This remarkable result demonstrates several points. First, the shift here is entirely due to the unconventional
pairing, which plays the role of an \emph{effective} SOC that couples $k$
and $\tau$. However, the spin here is the Nambu
pseudospin, which is intrinsic and unique for superconductors. Second, as a general advantage of the symmetry argument, as long as symmetry is preserved, the result does not depend on the details of the interface (see Fig.~\ref{fig: AR_USC_pwave}). Third,
the result in (\ref{eq:chiralp}) also applies for chiral pairings with
higher orbital moments ($|\chi|>1$), such as $d+id$ or $f+if$ pairings.

\begin{figure}[t]
\includegraphics[width=8cm]{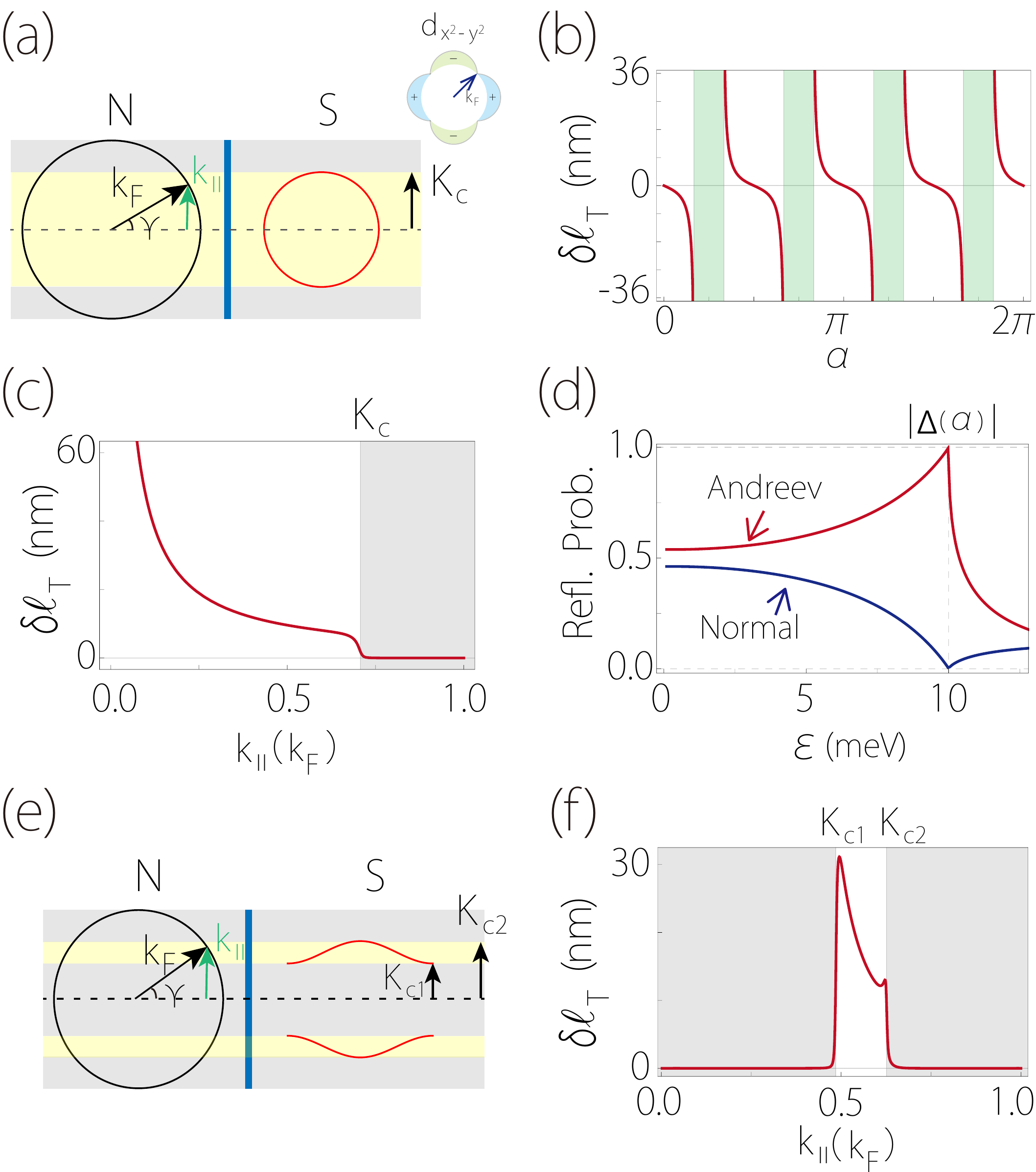}
\caption{
Results for $d_{x^2-y^2}$-wave pairing. (a-d) are for the S side with an ellipsoidal (closed) Fermi surface, and (e-f) are for S with a cylinder-like (open) Fermi surface.
(a) Schematic figure showing the Fermi surfaces of N and S. $K_c$ denotes the maximum magnitude of transverse wave-vector on the S Fermi surface.
(b) $\delta\ell_T$ versus  $\alpha$. The green shaded regions indicate the suppressed zones, in which  $\varepsilon>|\bm \Delta(\alpha)|$.
(c) $\delta\ell_T$ versus $k_\|$.
Corresponding to (a), $\delta\ell_T$ is suppressed  when $k_\|>K_c$, as denoted by the gray shaded region.
(d) Reflection probabilities versus $\varepsilon$ for normal and Andreev reflections.
(e) illustrates the case when the S Fermi surface is of open cylinder-like shape. $K_{c1}$ and $K_{c2}$ denote the lower and upper bounds for the transverse wave-vector on the S Fermi surface.
For such case, the qualitative features in (b) and (d) remain the same. The main difference is that the shift is now suppressed in regions except for $K_{c1}<k_\|<K_{c2}$, as shown in (f).
Figure adapted with permission from Ref.~\cite{YuSUTDPRL2018}.}\label{fig: AR_USC_dwave}
\end{figure}

Quantum scattering approach was adopted to study the transverse shift for other types of pairing in Table~\ref{T1}. An important case is for the $d_{x^2-y^2}$-wave pairing, with $\bm \Delta=\Delta_0\cos(2\phi_k)$. Here, because the pair potential is anisotropy in the $x$-$y$ plane, the transverse shift depends on the orientation of the plane of incidence. Hence, one needs to define a rotation angle $\alpha$ between the incident plane and the crystal $x$ axis.
It was found through calculation that
\begin{equation}\label{dwave}
\delta \ell_T \propto \sin(4\alpha)\Theta(|\Delta_{0}\cos2\alpha|-\varepsilon).
\end{equation}
The expression in Eq.~(\ref{dwave}) highlights the dependence on the rotation angle $\alpha$ and the excitation energy $\varepsilon$. The typical behavior is shown in Fig.~\ref{fig: AR_USC_dwave}.

One can observe the following key features for the shift. (i) The shift has a period of $\pi/2$ in $\alpha$, and it flips sign at multiples of $\pi/4$ [Fig.~\ref{fig: AR_USC_dwave}(b)]. (ii) The shift is sensitive to the gap magnitude. As indicated by the step function in Eq.~(\ref{dwave}), it is suppressed for excitation energies above the pairing gap at the incident wave vector. (iii) Particularly, due to the nodal structure of the gap, for a fixed excitation energy $\varepsilon$, there must appear multiple zones in $\alpha$ where $\delta\ell_T$ is suppressed [see Fig.~\ref{fig: AR_USC_dwave}(b)]. The center of each suppressed zone coincides with a node. (iv) The shift is also suppressed when $k_\|$ is away from the Fermi surface of the S side, as indicated in Fig.~\ref{fig: AR_USC_dwave}(c) and \ref{fig: AR_USC_dwave}(f), where we compare the results for a closed ellipsoidal Fermi surface and for an open cylinder-like Fermi surface. This can be understood by noticing that the effect of pair potential diminishes away from the Fermi surface.

These features encode rich information about the unconventional gap structure, including the $d$-wave symmetry [feature (i)], the gap magnitude profile [feature (ii)], and the node position [feature (iii)]. Feature (iv) also offers information on the geometry of the Fermi surface. Thus, by detecting the effect, one can extract important information about the unconventional superconductor.

Real unconventional superconductor
materials could have other complicated features, such as
multiple Fermi surfaces, multiple bands with different
pairing magnitudes, and possible interface bound
states~\cite{Tsuei2000,Sigrist2005,Mackenzie2003,Kallin2016}. How these features would affect the
anomalous shifts are interesting questions to explore. Nevertheless, the analysis in Ref.~\cite{YuSUTDPRL2018} suggested that a
nonzero shift is generally
expected, owing to the coupling between the Nambu
pseudospin and the orbital motion as generated by the unconventional pair potential. Although its detailed profile
requires more accurate material-specific modeling, it is likely that the key
features for the shift (as those listed in Table~\ref{T1}) are robust, since they
are determined by the overall characteristic associated with
the symmetry of unconventional pairings. This also helps to
distinguish the signal from the shift against random noises
such as from the impurities or interface roughness. Finally,
when the SOC effect is included, it can generate an
additional contribution to the shift. However, its dependence on the incident geometry
and the excitation energy will be distinct from that due to
the unconventional pairings.

\section{Transverse Shift in Crossed Andreev Reflection}

\begin{figure}[t]
	\includegraphics[width=8 cm]{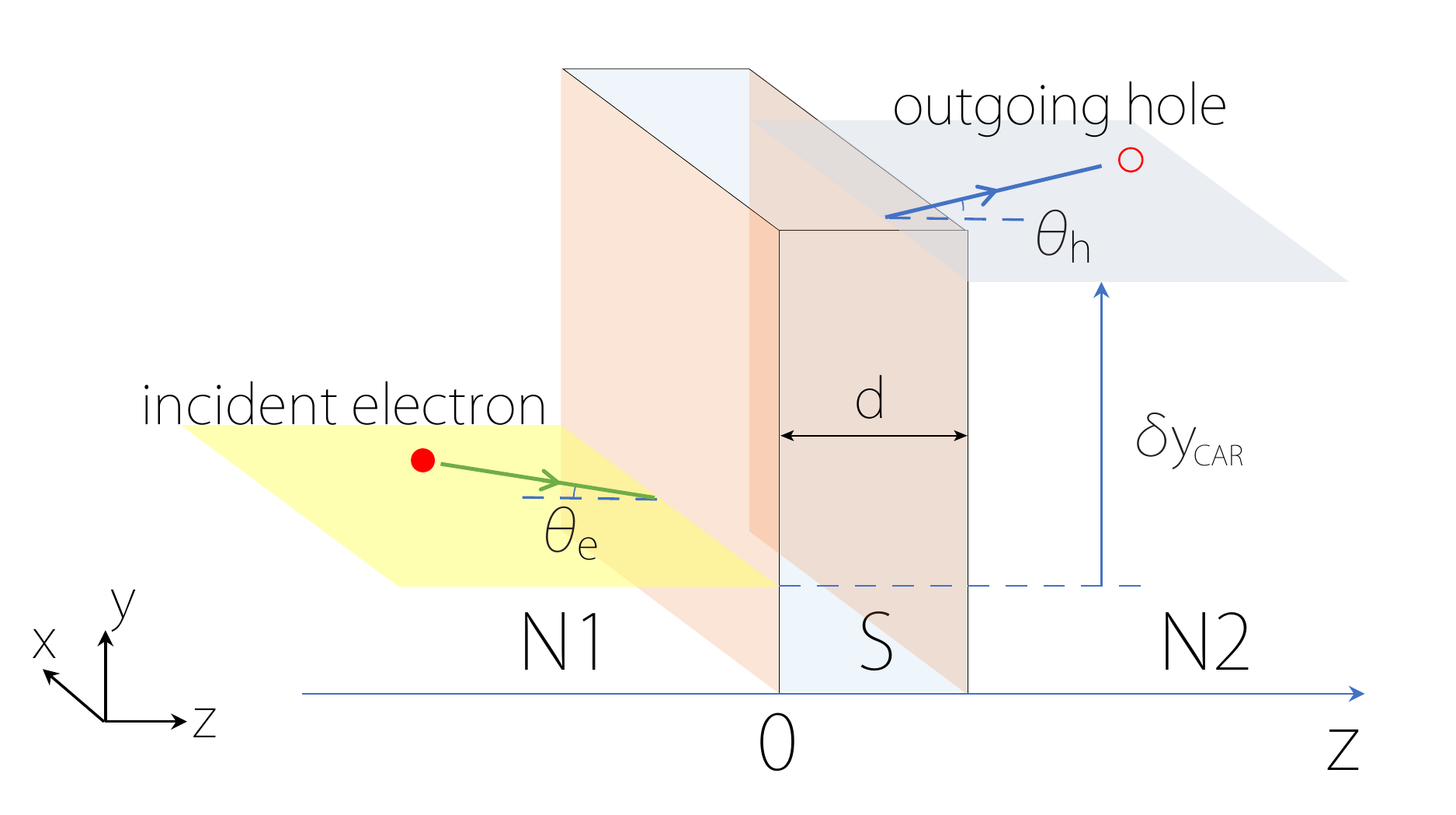}
	\caption{Schematic figure showing the process of CAR. In the hybrid NSN structure, an incident electron from terminal N1 is coherently scattered as an outgoing hole in terminal N2. There may exist a transverse shift $(\delta y_\mathrm{CAR})$ between the two scattering planes. Figure adapted with permission from Ref.~\cite{LiuCAR2018}.}
	\label{fig: ill_CAR}
\end{figure}

In 2018, Liu \emph{et al.}~\cite{LiuCAR2018} extended the study to the process of crossed Andreev reflection (CAR). CAR is a nonlocal version of the conventional Andreev reflection~\cite{Byers1995,Deutscher2000}. It appears in hybrid normal-superconductor-normal (NSN) structures, as schematically illustrated in Fig.~\ref{fig: ill_CAR}. When the thickness of the S layer is smaller than or comparable to the superconducting coherence length, an electron incident from the left N terminal can form a Cooper pair in S with another electron from the right N terminal, thereby coherently transmitting a hole into the right N terminal. The process has been successfully detected in experiment~\cite{Beckmann2004,Russo2005,Cadden-Zimansky2006}.

Using the quantum scattering approach and the symmetry argument, Liu \emph{et. al.}~\cite{LiuCAR2018} predicted that sizable transverse shift $\delta y_\text{CAR}$ can also exist in CAR. They considered systems where the N terminals have strong SOC described by similar models as in Ref.~\cite{LiuSUTDPRB2017} (including Weyl model and spin-orbit-coupled-metal model), and the S layer is of conventional $s$-wave superconductor. For such setups, the transverse shift is resulted from the SOC in the N layers. Compared with the local Andreev reflection studied in Ref.~\cite{LiuSUTDPRB2017}, a new ingredient here is that the two N terminals (hence the incident electron and the scattered hole) can be controlled independently. Particularly, one can use doped semiconductors as the two terminals, and make the left N terminal $n$-doped and the right N terminal $p$-doped, realizing a so-called $p$S$n$ junction~\cite{VeldhorstPRL2010}. For this kind of setup, one can minimize a competing transmission process---the elastic cotunneling, during which an incident electron directly tunnels through the structure. It was shown that the transverse shift still exists for the CAR holes, which may results in a measurable voltage signal, providing a new method for detecting CAR in experiment.

\section{Longitudinal Shift in Andreev Reflection}

For completeness, here we also briefly discuss the research on the longitudinal shift in Andreev reflection.
As we have mentioned in Sec.~II, the longitudinal shift is essentially a 2D effect, so it can be studied using 2D systems.
It was noticed that a previous theoretical work~\cite{Lee2013} studied this effect for a model based on a 2D electron gas, however, the shift was found to be absent. In 2018, Liu \emph{et al.}~\cite{LiuSUTD2018} showed that the result in Ref.~\cite{Lee2013} is actually a special limiting case. The longitudinal shift does exist for the general case, and it and can be quite sizable.

\begin{figure}[t]
\includegraphics[width=6cm]{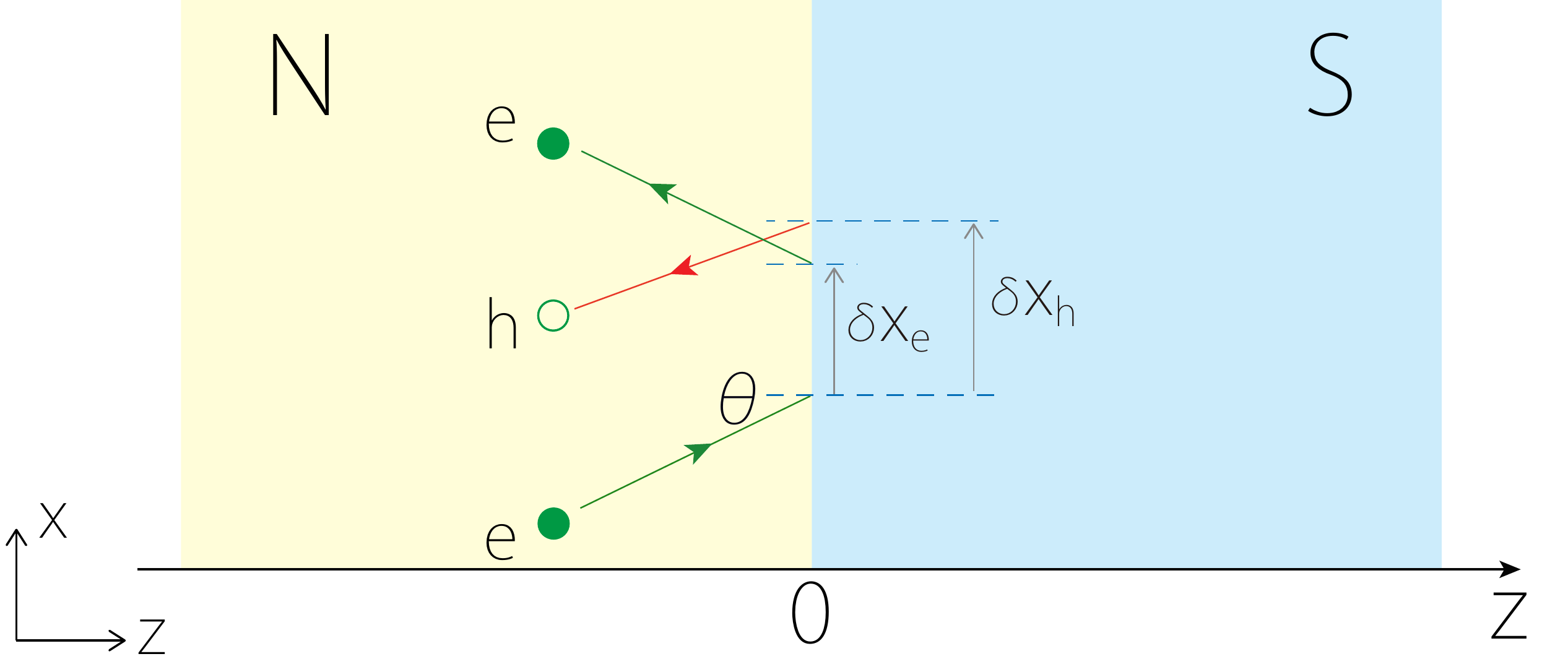}
\caption{Schematic figure showing the longitudinal shift in normal reflection ($\delta x_e$) and in Andreev reflection ($\delta x_h$) for an incident electron beam reflected from an NS interface. The solid and open circles indicate the electron and the hole, respectively. Figure adapted with permission from Ref.~\cite{LiuSUTD2018}.}
\label{fig11}
\end{figure}

Liu \emph{et al.}~\cite{LiuSUTD2018} studied two concrete examples. The first is a junction based on the simple 2D electron gas model, with
\begin{equation}
H_L=\frac{1}{2m}k^2,
\end{equation}
and
\begin{equation}
H_R=\frac{1}{2m}k^2-U_0.
\end{equation}
The system is assumed to be in the $x$-$z$ plane, as illustrated in Fig.~\ref{fig11}.

The calculation of the longitudinal shift via the quantum scattering approach is straightforward. It was found that
when the N side is heavily doped such that $E_F\gg U_0, \Delta_0, \varepsilon$, the Andreev reflection amplitude is given by a $k$-independent number
$
 r_h=e^{-i\beta}$.
Hence, in this regime, the longitudinal shift in Andreev reflection vanishes: $\delta x_h=0$. This recovers the result obtained in Ref.~\cite{Lee2013}, which assumed this regime.

\begin{figure}[t]
\includegraphics[width=8cm]{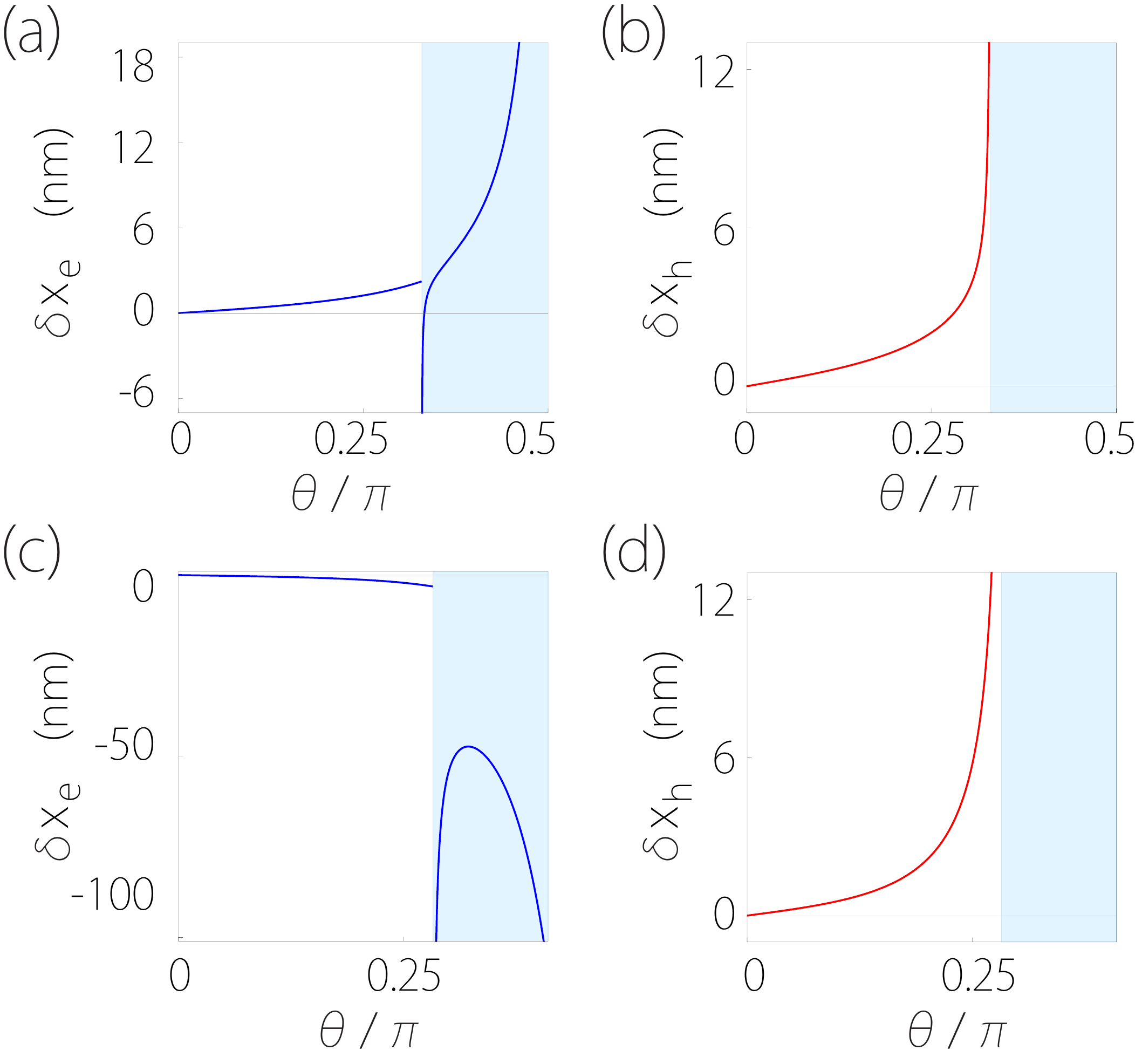}
\caption{Longitudinal shifts (a,c) in the normal reflection ($\delta y_e$), and (b,d) in the Andreev reflection ($\delta y_h$) versus the incident angle $\theta$ for the 2DEG/superconductor model. (a-d) are for small  $\varepsilon<\Delta_0$; while (c,d) are for $\varepsilon$ close to $\Delta_0$. The shaded region in each figure denotes the range with $|\theta|>\theta_c$, where Andreev reflection is not allowed.
Figure adapted with permission from Ref.~\cite{LiuSUTD2018}.}
\label{Fig_EGsft}
\end{figure}

However, outside of the above regime, when $E_F$ is not large, the shifts would generally be nonzero. Typical behavior of this shift is shown in Fig.~\ref{Fig_EGsft}. In addition, there may also be sizable shift for the normal reflection, as shown in Fig.~\ref{Fig_EGsft}(a),(c). It is interesting to note that while the shift in Andreev reflection stays positive as shown in Fig.~\ref{Fig_EGsft}(b),(d), the shift in normal reflection can be made either positive or negative, depending on the excitation energy. An explanation of this behavior is provided in Ref.~\cite{LiuSUTD2018}.

\begin{figure}[t]
\includegraphics[width=8cm]{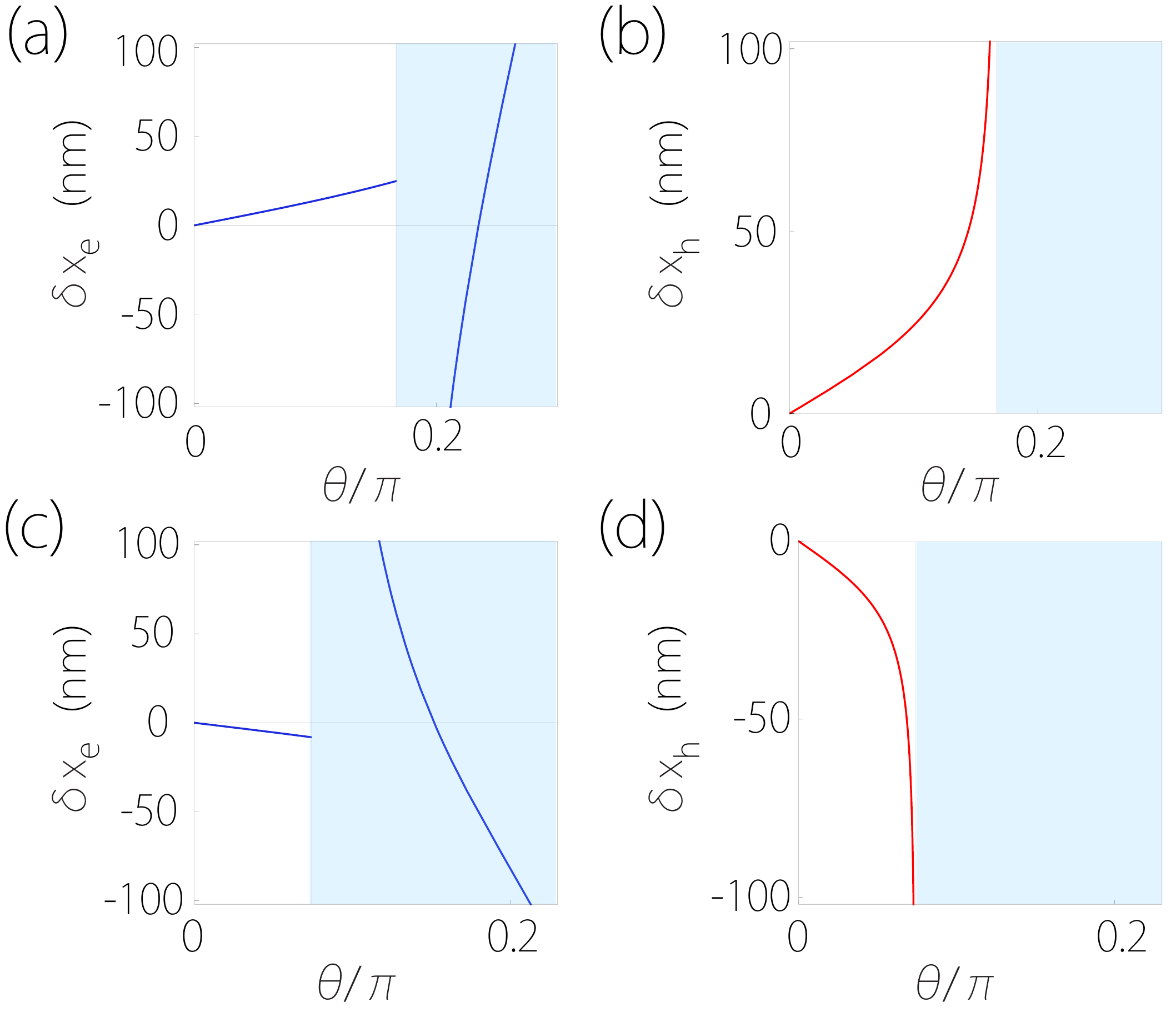}
\caption{Longitudinal shifts (a,c) in the normal reflection ($\delta {x_e}$), and (b,d) in the Andreev reflection ($\delta {x}_h$), as functions of the incident angle for the graphene/superconductor model.
(a,b) are for the case with $\varepsilon<E_F$; while 
(c,d) are for the case with $\varepsilon>E_F$. 
The shaded region in each figure denotes the region of $|\theta|>\theta_c$ where the Andreev reflection is not allowed.
Figure adapted with permission from Ref.~\cite{LiuSUTD2018}.}
\label{Fig_grasft}
\end{figure}

The second example is an NS junction based on graphene~\cite{Novoselov2004}. The graphene band structure has two Dirac cones (valleys) located at the two corner points $\pm K$ of the hexagonal Brillouin zone, which are connected by the time reversal symmetry~\cite{Neto2009}. In the BdG model,
\begin{eqnarray}\label{HG}
	H_L(\bm k,\tau)=v_F(\tau k_x\sigma_x+k_y\sigma_y),
\end{eqnarray}
where $\tau=\pm$ denotes the two valleys, $v_F$ is the Fermi velocity, the wave-vector is measured from the valley center, and $\sigma$'s are the Pauli matrices acting on $A/B$ sublattice degree of freedom. Note that $\mathcal TH_L(\bm k,\tau)\mathcal T^{-1}=H_L(-\bm k,-\tau)$, indicating that an incident electron in one valley is coupled to the hole in the other valley through the superconducting pair potential.
The S region is assumed to be described by the same Hamiltonian (\ref{HG}) (but with a nonzero pair potential and a potential energy offset). Physically, this may be realized by covering the graphene in the S region with a superconducting electrode, which induces a finite $\Delta$ by proximity effect. The potential energy offset $U$ may be adjusted by gate voltage or by doping. This model has been used by Beenakker~\cite{Beenakker2006} in discussing the special specular Andreev reflection in graphene.

\begin{figure}[t]
\includegraphics[width=8cm]{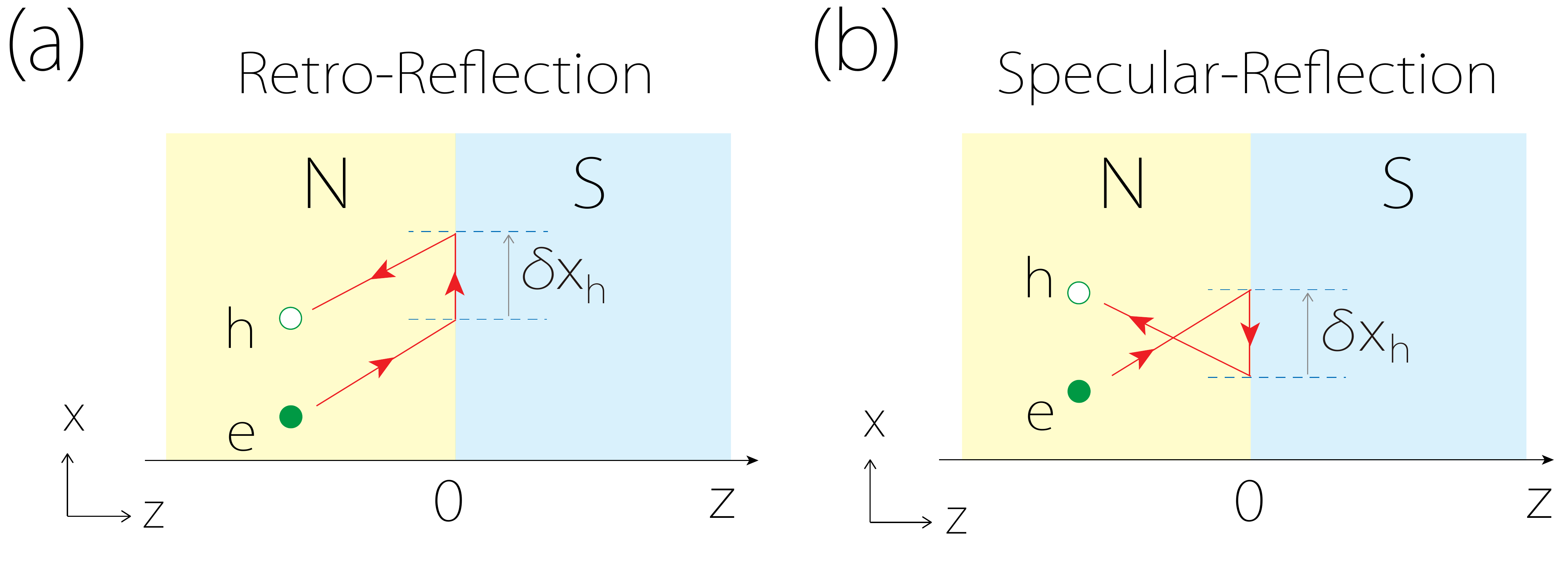}
\caption{Schematic figures for the shifts in the two types of Andreev reflection in the graphene/superconductor model. (a) is for
the retroreflection and (b) is for the specular reflection. Note that the shifts have opposite signs for the two cases. Figure adapted with permission from Ref.~\cite{LiuSUTD2018}.}
\label{Fig_refl}
\end{figure}

The calculation result showed that the longitudinal shift is enhanced by the additional pseudospin degree of freedom for graphene (see Fig.~\ref{Fig_grasft}). In addition, the shift in Andreev reflection can also be
made negative, and the sign is connected with whether the Andreev reflection is a retroreflection or a specular reflection, as illustrated in Fig.~\ref{Fig_refl}.

\section{Experiment}

The anomalous shifts in electronic systems have not been directly detected in experiment at the time of this review. Nevertheless, several possible experimental schemes have been put forward.

The most direct way is to produce a collimated electron beam to be scattered at the interface, and to detect the trajectory of the scattered beam. This can in principle be achieved by using local gates and collimators, as having been developed in the field of electron optics~\cite{Spector1990,Molenkamp1990,Dragoman1999}.

\begin{figure}[b]
\includegraphics[width=8cm]{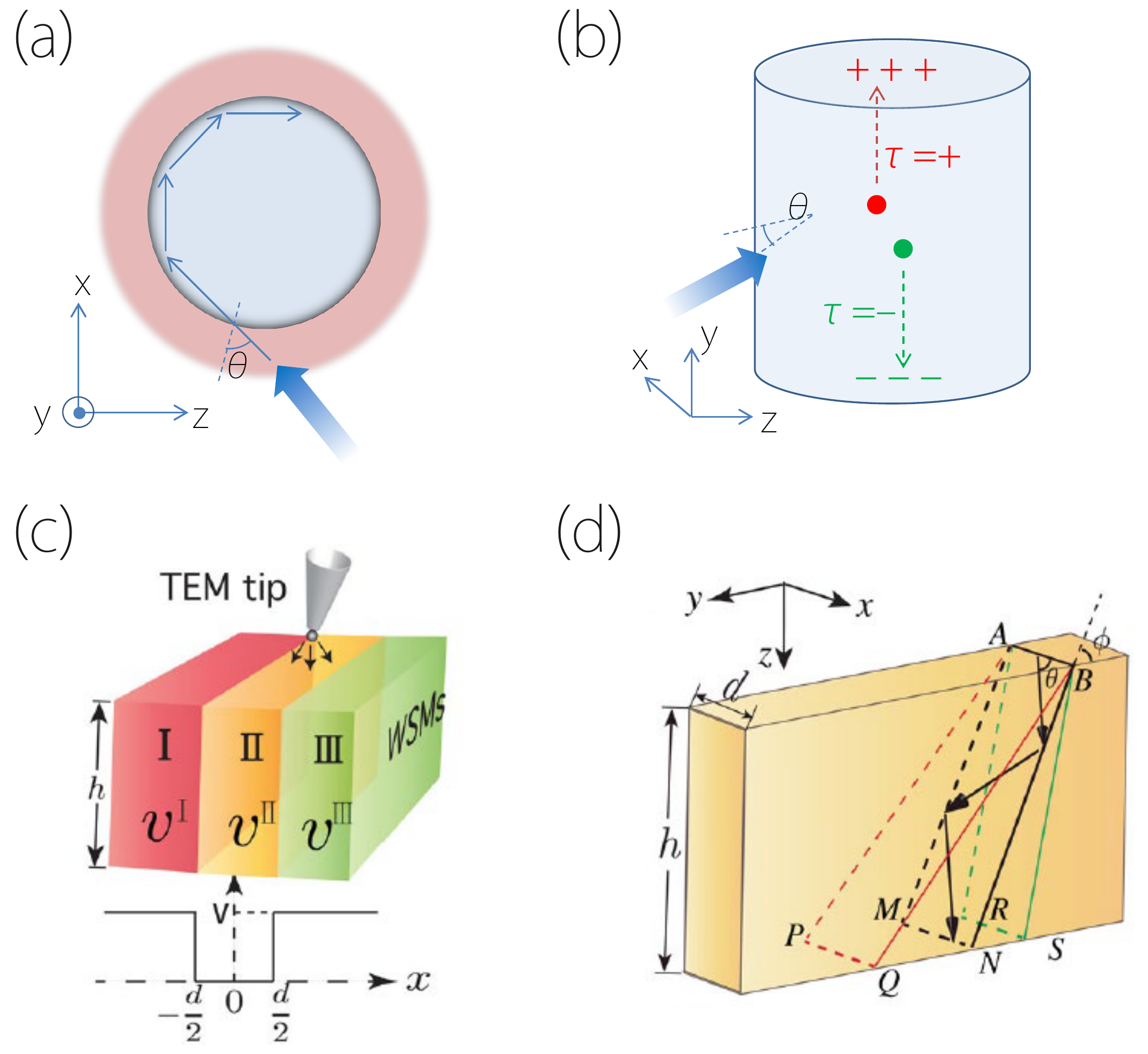}\caption{
(a) Top view of an electron undergoing multiple (total) reflections in a cylindric potential well of Weyl semimetals. (b) Side view of the enhanced chirality-dependent Hall effect in (a). (a,b) are adapted with permission from Ref.~\cite{YangPRL2015}.
(c) Schematic of a chirality splitter for Weyl fermions. Regions I, II, and III are three Weyl semimetal layers with different Fermi velocities. (d) Illustration of the wavepacket trajectory (the black arrow) of Weyl fermions in region II. The electrons are injected by a TEM tip.
(c,d) are adapted with permission from Ref.~\cite{JiangPRL2015}.}\label{fig: IF_dect1}
\end{figure}

\begin{figure}[tbh]
\includegraphics[width=8cm]{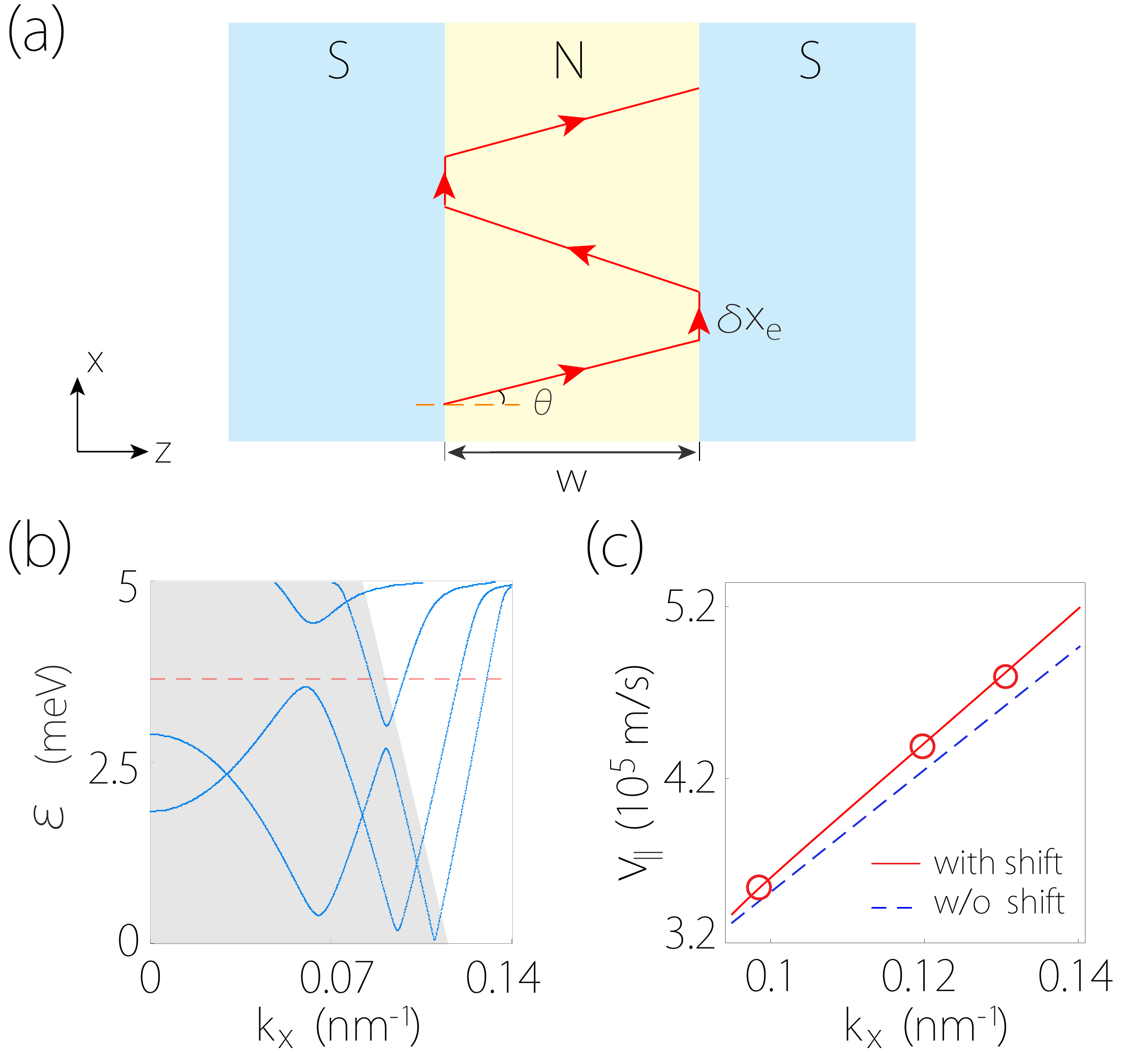}
\caption{(a) Schematic figure showing the trajectory for an electron confined in the SNS structure. Its propagation velocity along the $x$ direction is affected by the presence of the longitudinal shift. (b) Numerical results for the spectrum of the confined modes in the SNS junction. (c) Group velocities $v_\|$ for the confined modes at energy marked by the dashed line in (b). In (c), the data points are obtained from the numerical results in (b), the red solid (blue dashed) line is the estimation with (without) the $\delta x_e$ correction. Figure adapted with permission from Ref.~\cite{LiuSUTD2018}.}
\label{fig17}
\end{figure}

One possible way to enhance the overall effect is to design a structure such that the beam can undergo multiple times of scattering and the shifts can be accumulated. For example, the cylinder-shaped setup and the sandwich setup were proposed to enhance the transverse shift in normal reflection for Weyl electrons~\cite{JiangPRL2015,YangPRL2015} (see Fig.~\ref{fig: IF_dect1}). A 2D SNS waveguide was proposed to enhance the longitudinal shift in Andreev reflection~\cite{LiuSUTD2018} [see Fig.~\ref{fig17}(a)]. The repeated shifts lead to an anomalous velocity, which modifies the group velocity of the waveguide confined mode, as indicated in Fig.~\ref{fig17}(b) and \ref{fig17}(c).

\begin{figure}[b]
\includegraphics[width=8cm]{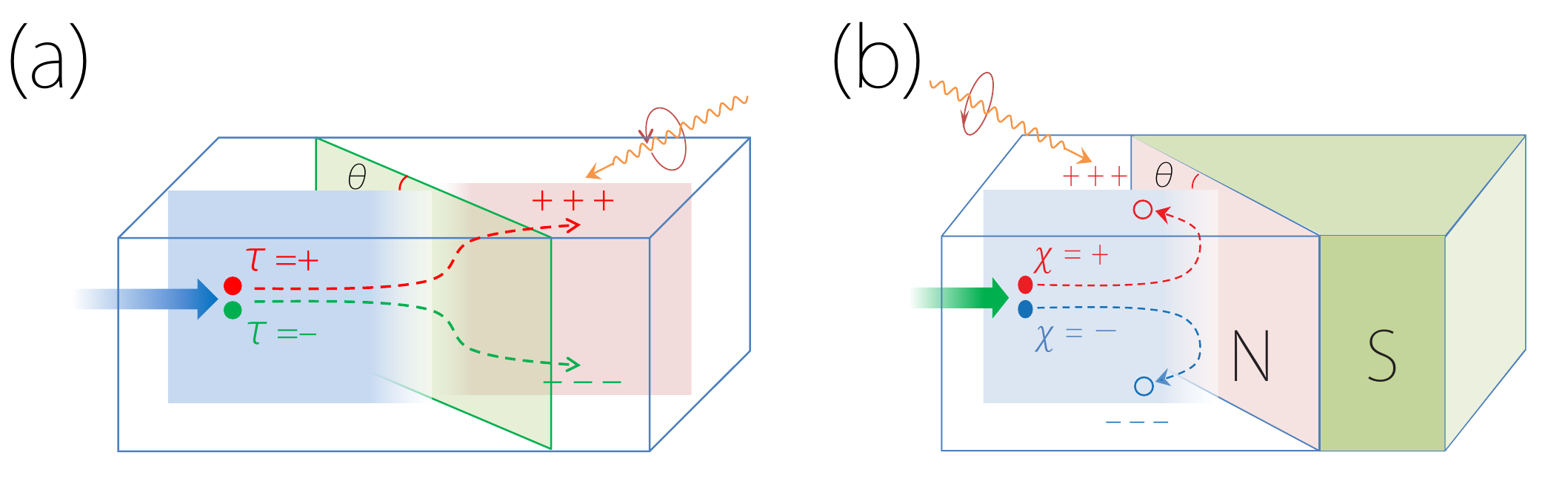}
\caption{(a) A Weyl semimetal junction structure. The transverse shift in transmission through the green colored interface induces a chirality accumulation at the top and bottom surfaces near the interface. (b) An NS junction with the N side being a Weyl semimetal. The transverse shift in Adreev reflection induces chirality accumulation for the reflected holes on the top and bottom surfaces near the interface. The chirality accumulation can be detected by the imbalanced absorbance for the circularly polarized lights. Figures adapted with permission from Refs.~\cite{YangPRL2015,LiuSUTD2018}.}
\label{fig18}
\end{figure}

\begin{figure}[t]
\includegraphics[width=8cm]{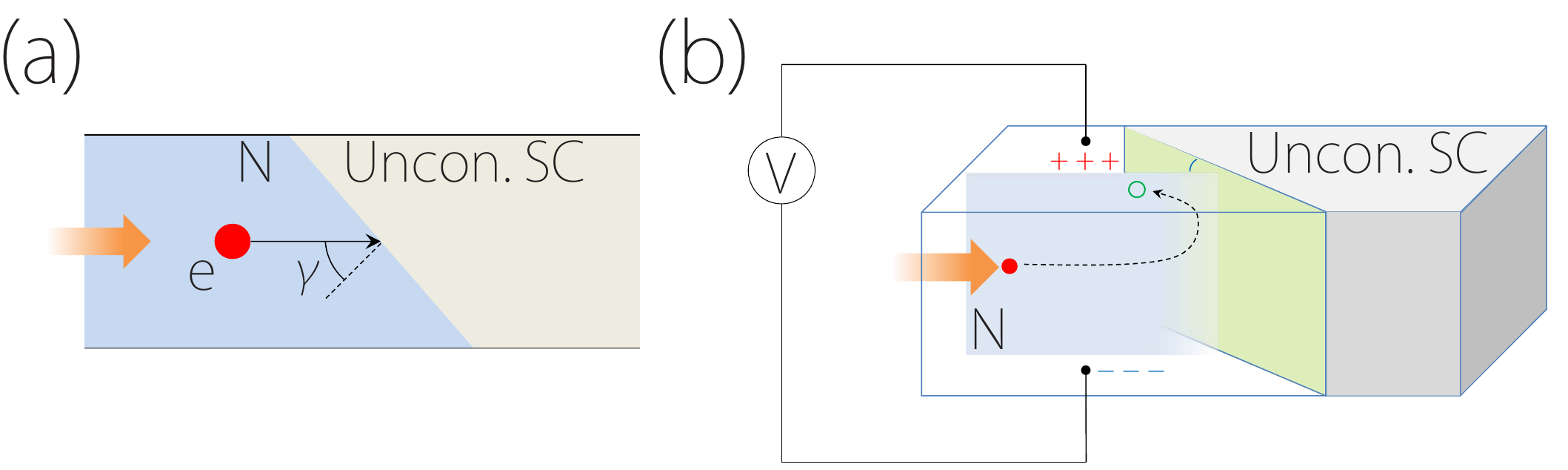}
\caption{ Schematic (a) top view and (b) side view of a possible NS junction with an unconventional superconductor. Electrons are driven to the interface with a finite average incident angle. The transverse shift induces a net surface charge accumulation near the junction on the N side, which can be detected as a voltage difference between top and bottom surfaces. The illustration is for the case when Andreev reflection dominates the interface scattering. Figure adapted with permission from Ref.~\cite{YuSUTDPRL2018}.}\label{fig: IF_dect2}
\end{figure}

A less challenging approach for detecting the transverse shift is to fabricating a junction with an interface tilted with respect to the average flow direction of the particles. Because the incident electrons hit the interface at a finite average incident angle, the average transverse shift for the outgoing particle will also be finite and has a definite sign. The transverse shift then leads to a net flow of the scattering particles, causing accumulation of the particles on the top (bottom) surface near the interface (see Figs.~\ref{fig18} and \ref{fig: IF_dect2}).
For Weyl semimetals, the shift depends on the chirality of the electron. If the electrons of opposite chiralities have equal population, there would be no net charge accumulation on the surface, but there is a surface chirality accumulation~\cite{YangPRL2015}. This can be detected by the imbalanced absorbance of the left and right circularly polarized light (see Fig.~\ref{fig18}). On the other hand, if the two populations are not equal, or for the junctions with unconventional superconductors (see Fig.~\ref{fig: IF_dect2}), a surface charge accumulation can be generated, which can be directly probed electrically as a voltage signal. In Ref.~\cite{YuSUTDPRL2018}, it was estimated that the voltage signal at an NS junction with unconventional superconductor can be up to mV magnitude, which can be readily detected with current experimental accuracy (on the order of nV).

\begin{figure}[b]
\includegraphics[width=7.cm]{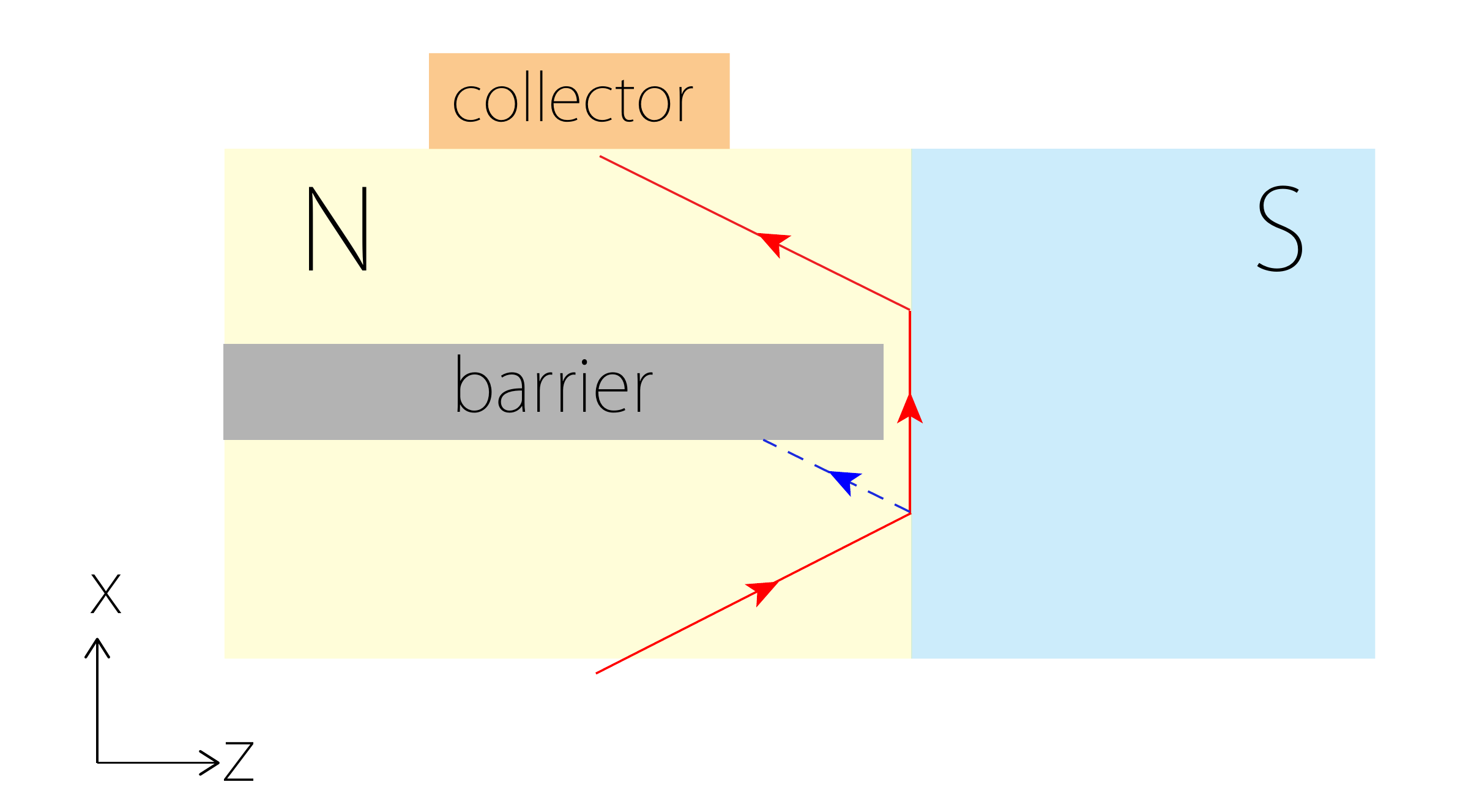}
\caption{ Schematic figure of a possible setup for detecting the longitudinal shift at an NS interface. Figure adapted with permission from Ref.~\cite{LiuSUTD2018}.
}
\label{Fig_detect}
\end{figure}

For the longitudinal shift in 2D systems, a possible experimental setup was proposed in Ref.~\cite{LiuSUTD2018}. As shown in Fig.~\ref{Fig_detect}, a collimated electron beam is incident onto the interface, and one tries to detect the reflected beam with the collector on the other side (see Fig.~\ref{Fig_detect}). The blue dashed line indicates the trajectory if there was no shift. We can engineer a barrier region (the gay colored one), e.g., by using local gating, such that the usual (dashed) trajectory is blocked. However, with the anomalous shift, the beam can circumvent the barrier region and follow the red path to be detected by the collector. Thus, the detection of the reflected beam at the collector will prove the existence of the shift.

It should be mentioned that real interfaces in the hybrid structures may have
roughness and imperfections. For interfaces that are too
disordered, it would be difficult to observe the anomalous shifts, because
the particles are strongly scattered and even the beam trajectory
is not well defined. However, one can expect that the shifts should be robust against weak disorders. This is because: (i) as shown from the studies, the shifts are not strongly oscillating functions with respect to the
incident angle, the energy, and etc., so they are not expected to
be averaged out under perturbations from weak roughness or
imperfections; (ii) typically, for doped semiconductors or semimetals, the Fermi wavelength can be made much larger than the atomic scale. Then the atomic-scale roughness or imperfections would have negligible effect on the shifts, since the particle simply does not
¡°see¡± them. Moreover, the advance in experimental technology
has made it possible to fabricate atomically sharp and clean
interfaces~\cite{ohtomo2004,liu2014}. Thus, the observation of the proposed effects should be within the reach of the available
experimental technique in the near future.

\section{Looking Forward}

The recent discovery of transverse shifts in normal electronic scattering and Andreev reflections has opened a new arena of physics research. The effect is intriguing and significant owing to the following aspects. First, the effect is intimately connected with geometric quantities such as the Berry curvatures. Second, the effect shows universal and robust features independent of the interface details when an emergent rotational symmetry exists, which is often the case in low-energy theories. Third, for the NS junctions, the rich behavior of the shift reflects the key features of the superconducting pair potential.

The above points indicate that the effect in the normal scattering can be utilized to probe the Berry curvatures of a material, an important task also connected to the study of topological materials. The shift in the Andreev reflection can provide a powerful new method to characterize the pair potentials of a superconductor. By mapping out the shift dependence on the incident geometry, one can in principle extract information of the symmetry of pairing, the gap magnitude, as well as the nodal structure.

On the theory side, we expect that the study will be extended to more types of junctions, e.g., with different types of topological materials or superconductors and with different junction configurations. This knowledge would be useful when one wants to use the effect to characterize different materials. The other direction is to apply this effect for designing functional devices. For example, the chirality dependence in the effect has led to the proposal of a chirality filter device in Refs.~\cite{JiangPRL2015,YangPRL2015}. To this end, more quantitative study of the resulting signal due to the shift is needed, which may be done by device modeling and numerical simulations.

On the experiment side, it is urgent to have the first demonstration of the effect. We expect that the longitudinal shift in Andreev reflection can be detected relatively easily, because the experimental techniques dealing with 2D systems are more advanced and mature. The detection of the voltage signals at the NS junction with unconventional superconductors is also an important task (as in Fig.~\ref{fig: IF_dect2}), for which the electrical detection of the voltage signal should be easier to perform.

\begin{acknowledgments}
We thank Xinxing Zhou and D. L. Deng for valuable discussions.
This work was supported by the Singapore Ministry of Education AcRF Tier 2 (MOE2017-T2-2-108).
\end{acknowledgments}

\bibliographystyle{apsrev4-1}
\bibliography{shift}

\end{document}